\documentclass[11pt,preprint]{aastex}
\begin{document}
\slugcomment{Accepted for publication in the Astrophysical Journal}
\title{Oxygen and Nitrogen in Leo A and GR 8}
\author{Liese van~Zee}
\affil{ Astronomy Department, Indiana University, 727 E 3rd St., Bloomington, IN 47405}
\email{vanzee@astro.indiana.edu}
\author{Evan D. Skillman}
\affil{Astronomy Department, University of Minnesota, 116 Church St. SE,
Minneapolis, MN 55455}
\email{skillman@astro.umn.edu}
\and
\author{Martha P. Haynes}
\affil{Center for Radiophysics and Space Research and
National Astronomy and Ionosphere Center,\altaffilmark{1} Cornell University, Ithaca, NY 14853}
\email{haynes@astro.cornell.edu}

\altaffiltext{1}{The National Astronomy and Ionosphere Center is
operated by Cornell University under a cooperative agreement with the
National Science Foundation.}

\begin{abstract}
We present elemental abundances for multiple HII regions in Leo~A and
GR~8 obtained from long slit optical spectroscopy of these two nearby 
low luminosity dwarf irregular galaxies.  As expected from their
luminosities, and in agreement with previous observations, the derived 
oxygen abundances are extremely low in both galaxies.  
High signal-to-noise ratio observations of a planetary nebula 
in Leo~A yield 12 + log(O/H) = 7.30 $\pm$ 0.05; ``semi-empirical'' calculations 
of the oxygen abundance in four HII regions in Leo~A indicate 
12 + log(O/H) = 7.38 $\pm$ 0.10.
These results confirm that Leo~A has one of the lowest ISM metal abundances
of known nearby galaxies.  
Based on results from two HII regions with high signal-to-noise measurements
of the weak [O~III] $\lambda$4363 line, the mean oxygen abundance of 
GR~8 is 12 + log(O/H) = 7.65 $\pm$ 0.06; using ``empirical'' and 
``semi-empirical'' methods, similar abundances are derived for 6 other 
GR~8 HII regions.  
Similar to previous results in other low metallicity galaxies, 
the mean log(N/O) $=$ $-$1.53 $\pm$ 0.09 for Leo~A and
$-$1.51 $\pm$ 0.07 for GR~8.
There is no evidence of significant variations in either O/H or
N/O in the HII regions.
The metallicity-luminosity relation for nearby (D $<$ 5 Mpc) dwarf 
irregular galaxies with measured oxygen abundances has a mean correlation
of 12 + log(O/H) = 5.67 $-$ 0.151 M$_{\rm B}$ with a dispersion in 
oxygen about the relationship of $\sigma$ $=$ 0.21.  
These observations confirm that gas-rich low luminosity galaxies have
extremely low elemental abundances in the ionized gas-phase of their 
interstellar media. 
Although Leo~A has one of the lowest metal abundances of known nearby 
galaxies, detection of tracers of an older stellar population (RR Lyrae 
variable stars, horizontal branch stars, a well populated red giant 
branch) indicate that it is not a newly formed galaxy as has been
proposed for some other similarly low metallicity star forming galaxies.
\end{abstract}

\keywords{galaxies: abundances --- galaxies: dwarf  --- galaxies: evolution  
--- galaxies: individual (Leo A, GR 8) --- galaxies: irregular}

\section{Introduction}

     Understanding the evolution of dwarf galaxies, and, in particular, 
the chemical evolution of dwarf galaxies, has implications for our
understanding of important processes in galaxy evolution, stellar
nucleosynthesis, and the enrichment of the intergalactic medium.
While a general relationship between galaxy luminosity and current 
metallicity has been well established for all galaxies, the 
fundamental underlying process (or processes) is still debated.
The star formation histories of dwarf galaxies are also far from well 
understood.  Is it possible for dwarf galaxies to form in the current
epoch, or are the identities of all dwarf galaxies established at
early times?  Can massive star formation result in the instantaneous
enrichment of the interstellar medium of a dwarf galaxy, or is the bulk
of the newly synthesized heavy elements released into a coronal gas
phase which must cool before becoming incorporated into the other 
phases of the ISM and eventually the next generation of stars?
By studying extreme dwarf irregular galaxies, i.e., the lowest 
luminosity, lowest metallicity and lowest mass star forming systems,
we can extend the baseline for comparative studies of galaxies and
test simple hypotheses of the formation of dwarf galaxies.  
This is particularly important for studying the chemical evolution
of dwarf galaxies, as the impact of a given episode of star formation
on a dwarf galaxy will be maximal for the lowest
luminosity and metallicity systems. 
In principle, this could lead to the largest deviations from the
luminosity-metallicity relationship, and this, in turn, may lead
to a better understanding of that relationship.

Gas phase abundances are one of the best measures of the intrinsic metallicity
of low mass galaxies.
Here, we present new optical spectroscopy of Leo~A and GR~8, two extremely 
low metallicity dwarf irregular galaxies in the Local Group
\citep[see][]{vdb00}.  Comparable nearby low luminosity galaxies
include DDO 210 \citep[no usable HII regions,]
[]{vHS97a}, Sag DIG \citep[empirical O/H only,][]{STM89,Se02,Le03c},
and Peg DIG (faint HII regions with no detectable [O~III] emission, \citet{SBK97}).
These (and other) observations indicate that it is often difficult to measure
the gas-phase abundance of low luminosity galaxies simply because there are 
few bright HII regions.  Thus, accurate oxygen abundance measurements
of multiple HII regions in Leo~A and GR~8 have the potential to
substantially further our understanding of
chemical enrichment and mixing of enriched material 
in low mass galaxies.

A summary of select global parameters of 
Leo~A and GR~8 is presented in Table \ref{tab:global}.
At a distance of only 690 kpc, Leo~A may be
the nearest low metallicity star forming galaxy.  Observations of the resolved
stellar population in Leo~A indicate an extremely low metallicity ($\sim$ 3\% of
solar) for even the youngest stars \citep{Te98}.  However, accurate gas
phase metallicities have been difficult to obtain for this system, in part
due to the low surface brightnesses of the faint HII regions.  The one exception is
the bright planetary nebula (PN) in the northeastern quadrant of the galaxy.
An oxygen abundance of $\sim$ 2\% of the solar value was derived from observations
of this planetary nebula \citep{SKH89}, but it was never certain if
this value was representative of the galaxy as a whole, or just the gas
ionized by the PN.  We thus set out to observe several of the HII regions in 
Leo~A in order to obtain a more representative estimate of the gas phase 
oxygen abundance.

GR~8 is slightly more luminous and metal--rich than Leo~A, but it is also
an extremely low metallicity system.  Previous oxygen abundance
measurements indicate that GR~8 has an oxygen abundance of approximately
5\% of the solar value \citep{SMTM88,MAM90}.  However, an unusually high N/O 
ratio derived for one of the HII regions suggested that there might be regions of
local nitrogen enrichment \citep{MAM90}.  We thus decided to observe several
HII regions in GR~8 to determine if there are regions of local nitrogen 
enrichment in this low mass galaxy.

This paper is organized as follows. In \S 2 we present spectroscopic observations of Leo~A and GR~8.
In \S 3 we present the nebular abundance calculations and discuss
direct and strong line abundance calibrations. In \S 4 we compare the
abundance determinations of these two low luminosity galaxies to values 
for other galaxies in the literature.
In \S 5 we demonstrate that even though Leo~A and GR~8 are among the lowest
metallicity systems known, observations of resolved stars in these galaxies 
indicate that star formation in these galaxies has occurred across 
the age of the universe.

\section{Observations}

\subsection{Optical Spectroscopy}
Long slit optical spectra of GR~8 and Leo~A were obtained with the
Double Spectrograph on the 5m Palomar\footnote{Observations at the 
Palomar Observatory were made as part of a continuing cooperative 
agreement between Cornell University and the California Institute of 
Technology.}  telescope on the nights of 18 and 19 February 1999.  
The observations were obtained in long--slit mode; the 2\arcmin~long--slit
was set to an aperture 2\arcsec~wide.  Complete
spectral coverage from 3600--7600 \AA~was obtained by
using a 5500 \AA~dichroic to split the light to the
two sides (blue and red) of the spectrograph.
The blue side was equipped with a 600 l/mm grating; the
red side was equipped with a 300 l/mm grating.
The effective spectral resolutions were well matched
between the two sides, with a resolution of
5.0 \AA~(1.72 \AA\ pix$^{-1}$) on the blue side and
7.9 \AA~(2.47 \AA\ pix$^{-1}$) on the red side.
Wavelength calibration was obtained by observations
of arc lamps obtained before and after the galaxy observations.
A hollow cathode (Fe and Ar) lamp was used to calibrate the blue
spectra; a combination of He, Ne, and Ar lamps were used to calibrate
the red spectra.

\begin{figure}
\epsscale{0.80}
\plotone{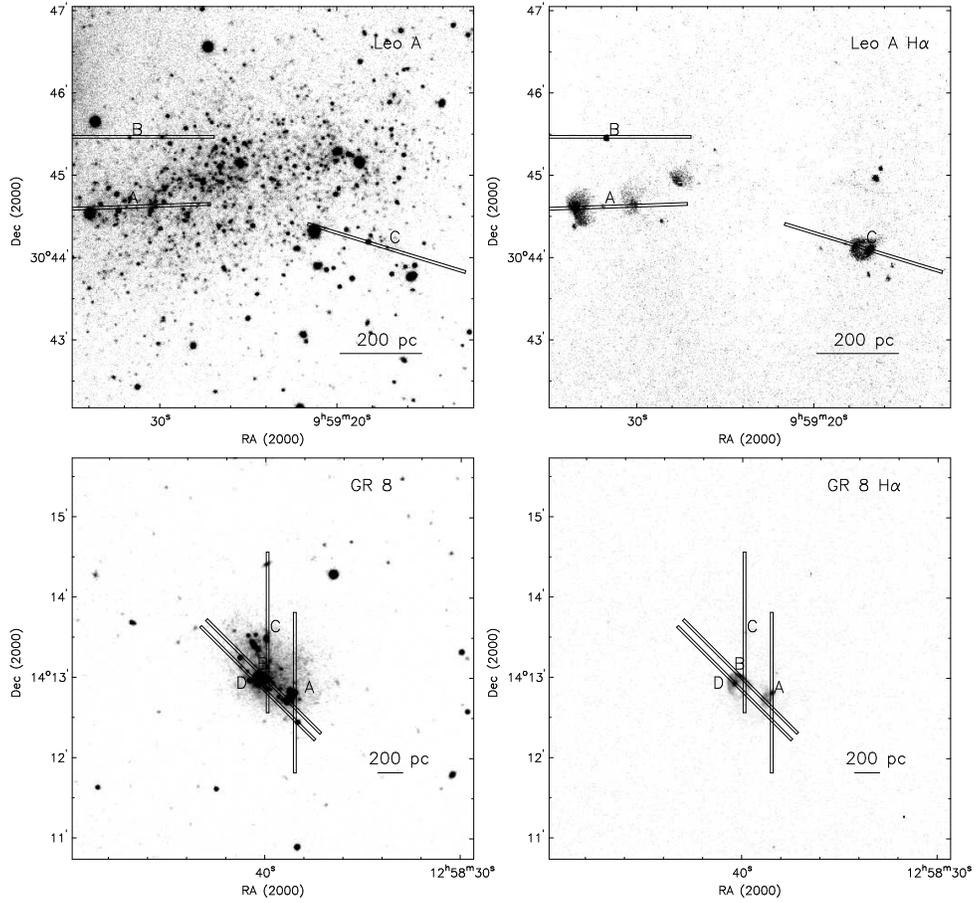}
\caption[]{
R-band and H$\alpha$ images of Leo~A and GR~8 with slit positions
marked.  The pointings and orientations were optimized to include multiple HII regions
while still maintaining a position angle close to the parallactic
angle at the time of the observation.
\label{fig:slits} }
\end{figure}

Stellar and HII region coordinates were obtained from H$\alpha$
images kindly provided by E. Tolstoy\footnote{
R-band and H$\alpha$ images of Leo~A were obtained on the 
KPNO 2.1m telescope as part of Hoessel \& Saha's KPNO 2.1m Cepheid 
search key program \citep{HSKD94}.}
 (Leo~A) and J. J. Salzer\footnote{
R-band and H$\alpha$ images of GR~8 were obtained on the KPNO 0.9m 
telescope as part of Salzer \& Westpfahl's survey of ionized gas in 
nearby galaxies \citep[cf.,][]{RSWR99}.} (GR~8).
H$\alpha$ imaging of Leo~A has previously been published by \citet{SHK91}
(but their field of view did not cover the westernmost HII regions) and
\citet{HHG93}.
H$\alpha$ imaging of GR~8 can also be found in \citet{HLK89} and references
therein.
Astrometric plate solutions for these images were calculated using the 
coordinates of bright stars in the APM catalog \citep{APM}, yielding 
positions accurate to within 1\arcsec.  Optimum slit positions were 
determined so that the slit could cross several HII regions during 
each observation while, at the same time, remaining close to the 
parallactic angle (Figure \ref{fig:slits}).  Due to the faintness of the HII 
regions, all observations were conducted via blind offsets from nearby
stars with typical offsets on the order of 30\arcsec~-- 80\arcsec.  
The observations were broken into a series of 20 minute exposures; typical 
observations were sets of 6 or 7 for HII regions in Leo~A and sets 
of 2 or 3 for HII regions in GR~8 (Table \ref{tab:obs}).

As can be seen in Figure \ref{fig:slits}, in Leo~A, the three slit 
positions include a mixture of HII region morphologies, while in
GR~8 the four slit positions include all major HII regions.
Previous Keck 10m observations of the two compact HII regions
in the northwest of Leo~A and the diffuse HII region located north of slit A
indicated that all of these HII regions have weak or non-detectable [O~II] and
[O~III] emission (H.\ A.\ Kobulnicky, private communication). 
Thus, the present observations concentrate on diffuse HII regions in Leo~A,
including the ring-like structure to the southwest (slit C), as well as
observations of the planetary nebula (slit B). 
The HII region observations also serendipitously included one background 
galaxy in each field.  Observations of the compact HII region south of 
slit C of Leo~A were obtained on 23 January 1999 under very cloudy 
conditions; while these observations are not discussed here since
the target HII region was only marginally detected 
(due to the clouds), this slit orientation included the background spiral 
galaxy at 09:59:15.85, 30:43:46.1 (J2000) with z = 0.1467 $\pm$ 0.0002.  
Finally, slit C of GR~8 included the background spiral 
galaxy at 12:58:39.95, 14:14:25.7 (J2000), with z = 0.0642 $\pm$ 0.0002.  

The spectra were reduced and analyzed with the IRAF\footnote{IRAF
is distributed by the National Optical Astronomy Observatories.} package.
Spectral reduction followed standard practice including bias subtraction, 
scattered light corrections, and flat fielding based on both twilight and dome flats.  
The 2--dimensional (2-d) images were rectified based on the arc lamp observations and the 
trace of stars at different positions along the slit.  Multiple exposures
were averaged together after the transformation.  The sky background was removed
from the 2-d images by fitting a low order polynomial along
each column of the spectra.  One dimensional (1-d) spectra of the HII regions 
were then extracted from the images.  
The 1-d spectra were corrected for atmospheric extinction and flux
calibrated based on observations of standard 
stars from the list of \citet{O90}; while the nights were not photometric, 
the relative line strengths should be robust.

\begin{figure}
\epsscale{0.80}
\plotone{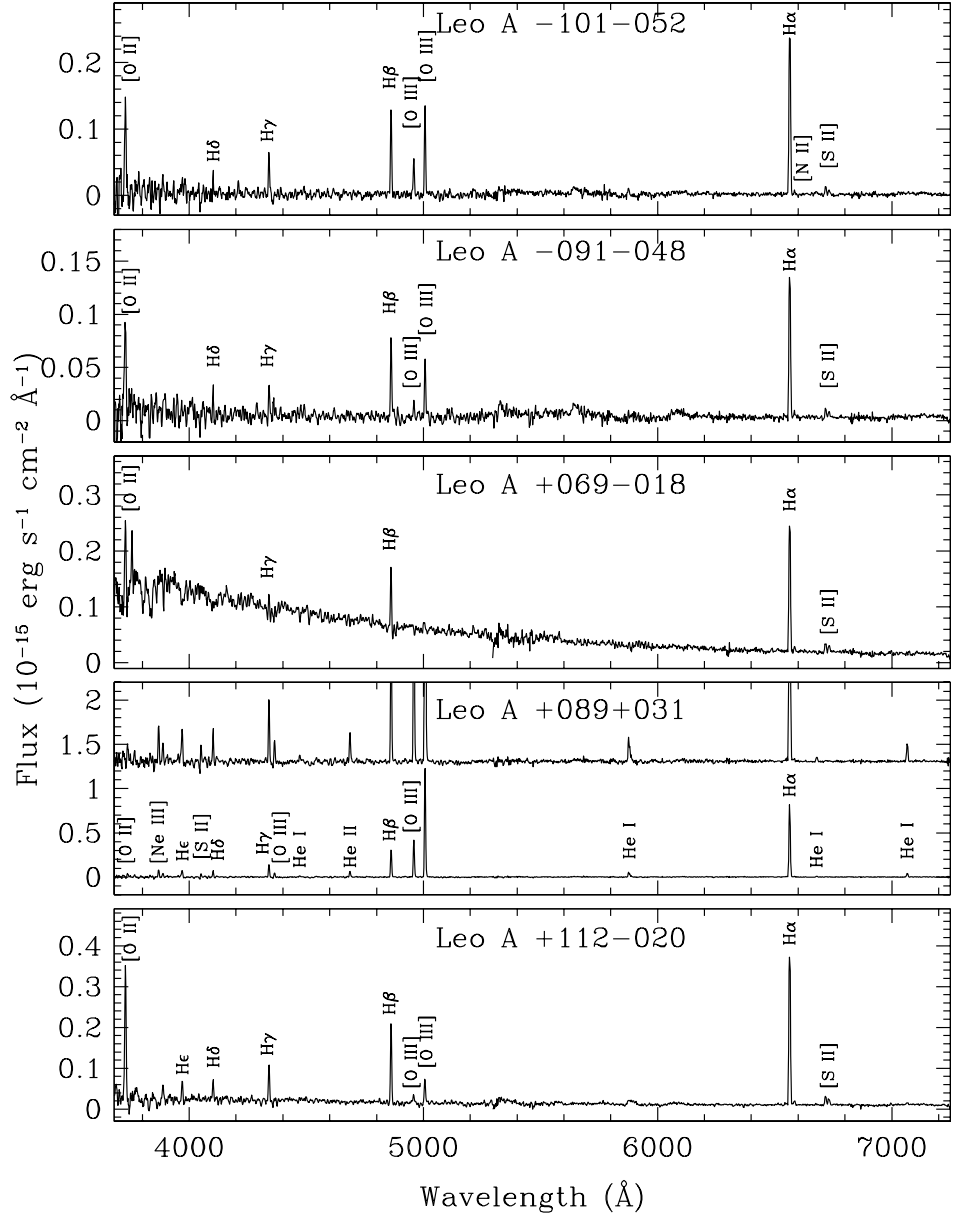}
\caption[]{
Optical spectra of HII regions in Leo~A.  Most of the HII regions
in Leo~A are low excitation; the one high excitation
spectrum (+089+031) is a planetary nebula. 
\label{fig:LeoA} }
\end{figure}

Representative spectra for Leo~A and GR~8 are shown in Figures \ref{fig:LeoA}
and \ref{fig:GR8}, respectively.  Note the excellent agreement in the 
continuum level in the blue and red spectra which confirms that the extraction
regions were well matched on the two cameras.  The HII regions in Leo~A are generally
low excitation, with weak [O~III] lines and moderate strength [O~II] lines.
The one exception is the spectrum of the planetary nebula (+089+031) which
has strong [O~III] and extremely weak [O~II] lines.  The weak 
[N~II] and [S~II] lines in these spectra clearly indicate that the 
HII regions in Leo~A have low elemental abundances.  The HII regions in 
GR~8 show higher excitation ([O~III]/[O~II]) than those in Leo~A;  
the [O~II] and [O~III] lines are strong in all of the GR~8 HII regions.  
However, similar to Leo~A, the weak [NII] and [S~II] lines indicate that 
the HII regions in GR~8 have low elemental abundances.

\subsection{Line Intensities}
The emission line strengths were measured in the 1-d spectra and then corrected
for underlying Balmer absorption and for reddening.  
Similar to the method described in \citet{vHS97a},
the strengths of the Balmer emission lines were used to estimate the amount
of reddening along the line of sight to each HII region using an
iterative technique.  The intrinsic Balmer line strengths were
interpolated from the tabulated values of \citet{HS87} for case B
recombination, assuming N$_e$ = 100 cm$^{-3}$ and T$_e$ = 15000 K.
Assuming a value of $R = A_V/E_{B-V} = 3.1$, the Galactic reddening
law of \citet{S79} as parameterized by \citet{H83} was adopted to
derive the reddening function normalized at H$\beta$.
For those HII regions with detected [O~III] $\lambda$4363, the
temperature was then recalculated from the corrected line strengths
and a new reddening coefficient was produced.
An underlying Balmer absorption with an equivalent
width of 2 \AA~was assumed in the few instances where the reddening coefficient
was significantly different when derived from the observed line ratios of
H$\alpha$/H$\beta$ and H$\gamma$/H$\beta$.
The reddening coefficients, c$_{H\beta}$, derived
for each HII region are listed in Tables \ref{tab:LeoA} and \ref{tab:GR8}.
These can be compared to the values of the foreground Galactic extinction
derived by \citet{SFD98} of E(B-V) $=$ 0.021 for Leo~A and 0.026 for 
GR~8 which correspond to c(H$\beta$) $=$ 0.03 and 0.04 respectively.
These very low values of foreground reddening are less than the errors
on all of our reddening measurements. 
In several instances, the derived reddening coefficients were slightly 
negative (which is not physical) and in these cases zero reddening was 
assumed.

The measured intensities of emission lines for the HII regions in Leo~A
and GR~8 are tabulated in Tables \ref{tab:LeoA} and \ref{tab:GR8}, respectively.
The first two columns in these tables give the ionic species and rest wavelength
of the transition.  The third and subsequent columns list the
extinction corrected line intensity relative to H$\beta$ for each
detected transition.  The H$\beta$ flux
and equivalent width for each object are listed at the bottom of
the tables; since the nights were non-photometric, the H$\beta$ fluxes
should only be taken as indicative values.  
The error associated with each relative line intensity
was determined by taking into account the Poisson noise in the line, the
error associated with the sensitivity function, the contributions of
the Poisson noise in the continuum, read noise, sky noise, and flat
fielding or flux  calibration errors, the error in setting the
continuum level, assumed to be at the 10\% level,
and the error in the reddening coefficient.  Note that although the spectral
coverage of the red camera allows for the measurement of the [O~II]
$\lambda\lambda$7320,7330 lines, these lines were not detected in any
of the spectra due to a combination of intrinsic faintness and strong
telluric emission.  The spectral coverage of the red camera did not
extend to the [S~III] $\lambda\lambda$9069,9523 lines.

\section{Nebular Abundances}

   In this section we will derive absolute and relative abundances from
the emission line spectra.  In order to avoid confusion, we start with 
an overview of some nomenclature.
In those cases where we have detected the temperature sensitive [O~III] 
$\lambda$4363 line, we can calculate abundances following the methods 
described in \citet{O89}, and, following \citet{D90} we will refer to 
this as the ``direct'' method.  In those cases where [O~III] $\lambda$4363 
is not observed, but [O~III] $\lambda$5007 and [O~II] $\lambda$3737 are 
observed, so-called ``empirical'' methods can be used to infer oxygen 
abundances \citep[e.g.,][]{PEBCS79,M91}. 
\citet{vHS97a} has introduced the ``semi-empirical'' method, which 
uses the oxygen abundance and ionization parameter implied from 
photoionization models to infer a consistent electron temperature.
That electron temperature can then be used to derive relative 
abundances with some degree of accuracy since the relative 
abundances have a much lower sensitivity to the electron 
temperature than the absolute abundances.
 
\subsection{Direct Abundance Measurements}

Direct calculation of the oxygen abundance from the observed [O~II] and [O~III] 
emission line strengths is possible if both the electron temperature (T$_e$) and 
electron density (N$_e$) are known.  The procedure to calculate T$_e$ and N$_e$ is 
well described in \citet{O89} and will not be repeated here.
A version of the FIVEL program of \citet{DDH87} 
was used to compute T$_e$ and N$_e$ from the reddening--corrected 
[O~III] and [S~II] line ratios, respectively.  In all cases, the
observed [S~II] line ratios indicated that the HII regions were within
the low density limit (I($\lambda$6716)/I($\lambda$6731) $>$ 1.35).
Thus, an N$_e$ of 100 cm$^{-3}$ was adopted for all HII regions.  Unfortunately,
calculating the electron temperature proved to be more difficult since
the weak [O~III] $\lambda$4363 line was often contaminated by the
nearby Hg $\lambda$4358 night sky line.  In Leo~A, [O~III] $\lambda$4363 
was detected in only one nebula, which is almost certainly a planetary
nebula.  While a hint of the [O~III] $\lambda$4363 line
is visible in several of the GR~8 HII regions (Figure \ref{fig:GR8}), it
was only detected solidly in two.  
Once the electron temperature of the O$^{++}$ ionization zone is derived,
the electron temperature of the O$^+$ zone is estimated using the approximation
given by \citet{PSTE92}.  The emissivity coefficients for all of the 
detected emission lines were then calculated using the FIVEL code \citep{DDH87}.

For all atoms other than oxygen, the derivation of atomic
abundances requires the use of ionization correction factors (ICFs) to account
for the fraction of each atomic species which is in an unobserved ionization state.
To estimate the nitrogen abundance, we assume that N/O = N$^+$/O$^+$  \citep{PC69},
which is probably accurate to $\pm$20\% for O/H $<$ 25\% solar \citep{G90}.
To estimate the neon abundance, we assume that Ne/O =  Ne$^{++}$/O$^{++}$ 
\citep{PC69}.
To determine the sulfur and argon abundances, we adopt an
ICF from published HII region models to correct for the unobserved S$^{+3}$
and Ar$^{+4}$ states \citep[e.g.,][]{TIL95}. 
In HII regions where S$^{+2}$  $\lambda 6312$ was not detected, we do not
calculate a sulfur abundance as the ICF becomes too uncertain \citep{G89}.

\subsubsection{Leo A +089+031 (PN)}
The highest surface brightness feature in the H$\alpha$ image of Leo~A
is the bright planetary nebula in the northeastern part of the galaxy.
Note that the H$\alpha$/H$\beta$ ratio is close to the theoretical value,
showing no evidence of the anomaly in the observations of \citet{SKH89}. 
The weak [O~III] $\lambda$4363 line is clearly detected in the high
signal--to--noise spectrum of this region (Figure \ref{fig:LeoA}).
However, it should be noted that [O~II] $\lambda$3727 was not detected
in this spectrum, and thus, in principle, the oxygen abundance derived
from the [O~III] emission (12 + log(O/H) = 7.30 $\pm$ 0.05) is a lower 
limit (Table \ref{tab:new}).  However, an upper limit on the 
[O~II] $\lambda$3727 line, listed in Table \ref{tab:LeoA}, corresponds 
to an O$^+$/H abundance of less than 4\% of the O$^{++}$/H abundance,
which is significantly smaller than the error on the total abundance.
Thus, we will treat this number as the best estimate of the oxygen 
abundance, noting that this result is similar to that derived previously
for the PN \citep{SKH89} in this low luminosity galaxy.
Note that, unfortunately, it is not possible to measure the N/O, S/O, and 
Ar/O ratios directly in this object because the [O~II] $\lambda$3727 was 
not detected and that is required for the ionization correction schemes 
for these elements.

\subsubsection{GR 8 --019--019 and +008--011}

Two of the HII regions in GR~8 have solid [O~III] $\lambda$4363 detections.
Following the procedure outlined above, the oxygen, nitrogen, neon, sulfur,
and argon abundances were calculated for GR~8 --019--019 and +008--011.
The electron temperatures derived from the [O~III] line ratios are listed
in Table \ref{tab:new}.  
The ionic abundances relative to hydrogen were computed using the data
listed in Table \ref{tab:GR8}; if more than one line was observed for
a given ion, the final ionic abundances were calculated to be the weighted
averages of the values for the different lines.  The nebular abundances
were then calculated based on the ionic abundances and the ICFs as described
above.  The derived 12 + log(O/H), log(N/O), log(Ne/O), log(S/O), and log(Ar/O)
for these HII regions are listed in Table \ref{tab:abund}.
The mean oxygen abundance for these two HII
regions is 12 + log(O/H) = 7.65 $\pm$ 0.06, in excellent agreement with previous 
direct oxygen abundance measurements for GR~8 
\citep[12 + log(O/H) = 7.48 $\pm$ 0.14; 7.68,][]{SMTM88, MAM90}.  

\subsection{Strong Line Oxygen Abundance Measurements}

As first discussed by \citet{PEBCS79}, the oxygen abundance
of an HII region can be estimated from the strong line ratio,
R$_{\rm 23} = $ ([O~II] + [O~III])/H$\beta$, since this parameter
varies smoothly as a function of stellar effective temperature and
oxygen abundance.  Furthermore, while this ratio is double valued,
the degeneracy of the R$_{\rm 23}$ relation can be
broken by examining the relative strengths of [N~II] and [O~II].
If the nitrogen lines are extremely weak, as is the case
for both Leo~A and GR~8 (Figures \ref{fig:LeoA} and \ref{fig:GR8}),
only the lower oxygen abundance branch needs to be considered.
However, an additional spread in the estimated oxygen abundance
for a given R$_{\rm 23}$ is introduced by the geometry
of the HII region \citep[e.g.,][]{S89, M91}; geometric effects
can be represented by the average ionization parameter, \=U,
the ratio of ionizing photon density to particle density.
This second parameter can be traced by the ratio of the abundance
of atoms in different ionization states.
Thus, both the sum of the oxygen lines (R$_{\rm 23}$) and the
ratio of [O~III] to [O~II] need to be considered before an
oxygen abundance can be determined from the strong lines.
For example, the low abundance branch 
of the model grid \citet{M91} generated for an IMF with an
upper mass limit of 60 M$_{\odot}$ is shown in Figure \ref{fig:r23}.
In this model (and other similar theoretical abundance calibrations, e.g.,
Kewley \& Dopita 2002), the R$_{\rm 23}$ value and [OIII]/[OII] ratio are 
used to derive an empirical estimate of the oxygen abundance.

As illustrated in Figure \ref{fig:r23},
the HII regions in GR~8 cluster around an oxygen abundance
of 12 + log(O/H) = 7.8 in the \citet{M91} calibration, indicating
that all of these HII regions have a similar oxygen abundance.
However, there is a troubling discrepancy between the oxygen abundances 
derived from the \citet{M91} calibration and those obtained by direct
calculation based on measured T$_{\rm e}$ and individual
line strengths \citep[e.g.,][]{P00,PMD05,vH04}.  In general, 
the \citet{M91} calibration appears to yield abundances that are 
systematically 0.1-0.2 dex higher than direct calculations;
this discrepancy is also found for the HII regions in GR~8
(direct abundance calculations yield 12 + log(O/H) = 
7.65 $\pm$ 0.06, Section 3.1.2).

An alternative empirical calibration scheme was proposed be \citet{P00} 
based on HII regions where [O III] $\lambda$4363 has been detected (the p-method).
The \citet{P00} relations for 12 + log(O/H) between 7.4 and
8.2 are shown in Figure \ref{fig:r23}.
In the high ionization regime (\=U $\sim$ 0.01 -- 0.1), 
the shapes of the McGaugh curves and the Pilyugin relations 
are reasonably similar, but the Pilyugin relations give oxygen 
abundances $\sim$ 0.12-0.15 dex lower for the same R$_{\rm 23}$ value.  
However, as discussed in \citet{SCM03} and \citet{vH04},
and is also clear from Figure \ref{fig:r23}, the empirical relations 
of \citet{P00} are only valid in this high ionization regime, 
where there were sufficient T$_{\rm e}$ measurements in the literature 
to enable such a calibration.  In the low ionization regime,
the regime of most of the HII regions in GR~8 and Leo~A,
the empirical relations of \citet{P00} clearly deviate from
results of photoionization models and physical intuition.  Indeed,
\citet{vH04} demonstrate that the empirical abundances derived
from the p-method have a clear systematic offset that correlates
with ionization parameter.  Thus, the empirical oxygen 
abundances listed in Table \ref{tab:abund} 
for the HII regions in Leo~A and GR~8 are based on the \citet{M91}
model grid, despite the inherent uncertainties and possible
systematic errors.  The empirical abundances of \citet{P00}
will not be discussed further.

While the GR~8 HII regions clearly outline a common locus in 
Figure \ref{fig:r23}, the HII regions in Leo~A appear to be more 
widely scattered in ionization parameter and, perhaps, oxygen abundance.  
Two of the observed HII regions have extremely low ionization parameters 
(Leo~A +069--018 and +112--020); both are extremely diffuse HII regions 
located in the eastern half of the galaxy.  For +069--018, 
[O~III] $\lambda$5007 is barely detected, indicating that this HII region 
has an extremely low ionization parameter.  Similar low excitation spectra 
were seen in the faint HII region of another Local Group galaxy, 
Peg DIG \citep{SBK97}.  For Peg DIG, the morphology (compact) and 
spectrum of the HII region is consistent with an ionizing flux of a 
single B0 star.  In Leo~A, the Balmer line flux that is observed corresponds to
ionization by B0 - O9.5 ZAMS stars for each of the 4 HII regions.  
It is thus possible that the IMF is truncated (or incompletely sampled)
or that these HII regions are evolved (aged).
However, unlike Peg DIG, the low excitation HII regions in Leo~A are 
diffuse, amorphous structures; thus, it is also possible that the long 
slit observations do not sample fully the different ionization zones of 
these HII regions.  In particular, if the observations are missing a 
significant fraction of the O$^{++}$ ionization zone, these HII regions 
may be displaced significantly in Figure \ref{fig:r23}; with additional
[O~III], their location would shift significantly upwards and slightly 
to the right, possibly moving them into the same metallicity regime as 
the other two HII regions.  

Nonetheless, since the Balmer line fluxes of the HII regions in Leo~A are
consistent with radiation from lower mass, cooler O and B stars, 
we examine further the effect of aging in the abundance diagnostic diagram.
In Figure \ref{fig:grid} we place the low metallicity models (12 + log(O/H) = 7.33)
of \citet{SL96} in the diagnostic grid of Figure \ref{fig:r23}.
As can be seen in Figure \ref{fig:grid}, aging of HII regions can 
introduce significant scatter in the abundance
diagnostic diagram as both the ionization parameter
and shape of the ionizing spectrum evolve \citep[e.g.,][]{SL96,O97},
Thus, once one allows for a large range of ionizing spectra and 
ionization parameter, HII regions of a single oxygen abundance 
can populate a large range of positions 
in this diagnostic diagram.  This is further emphasized in the 
bottom panel of Figure \ref{fig:grid} where the positions of 
theoretical HII regions with the same oxygen 
abundance as the models in the upper part of the Figure are shown 
with varying values of the electron temperature and ionic fraction.  

While the results of Figure \ref{fig:grid} may appear to suggest
that strong line ratios alone are not adequate 
to determine an empirical oxygen abundance, it is important to recall
that the empirical scheme of \citet{M91} (and others) relies on a 
limited range of radiation field and a tight coupling between 
electron temperature and ionic fraction.
Since both the spectrum of the ionizing
radiation and the ionization parameter affect the ionic fraction,
if the ionizing radiation is not held fixed (e.g., zero-age main
sequence stars with a fully populated IMF) then there is a larger
scatter for a given abundance. 
However, these parameters are likely a reasonable approximation for most
high surface brightness HII regions, and thus the photo-ionization
models do yield reasonable estimates of the oxygen abundance in
most instances.  Rather, the issue of varying the ionizing radiation
only becomes significant for low excitation HII regions.  
In fact, the tight correspondance between 
the iso-metallicity contours and the location of the HII regions 
in GR~8 in Figure \ref{fig:r23} suggests that log(O32) $\ge$
$-$0.4 could serve as a guideline for a lower limit of the
utility of the R23 calibration.

In any event, whether the HII regions in Leo~A are evolved, were formed with 
a truncated IMF, or were incompletely sampled by the observations, 
it is likely that the model grid shown in Figure \ref{fig:r23} is 
not optimal for their analysis and interpretation.  
Thus, we caution that it is premature to speculate
on possible oxygen abundance variations in Leo~A; in particular, 
the planetary nebula has an oxygen abundance similar to the estimated 
abundances of the other two (higher excitation) HII regions 
(12 + log(O/H) $\sim$ 7.46).  In fact, while we cannot rule out oxygen
abundance variations, all of the HII region spectra in Leo~A
are consistent with the oxygen abundance determined in the 
bright planetary nebula (see Section 4.1). 

\subsection{Semi--Empirical Relative Abundance Measurements}

\subsubsection{HII Regions in Leo A}
The [O~III] $\lambda$4363 line was not detected in any of the 
other Leo~A HII region spectra.
Thus we adopt a ``typical''
electron temperature of 15000 $\pm$ 2500 K to enable analysis of
nitrogen, neon, sulfur, and argon abundances.   While we have
no representative HII regions with measured T$_{\rm e}$ in
this galaxy to use as a basis for this value, there is a
general anti--correlation between electron temperature and
oxygen abundance because low metallicity gas is cooled
less efficiently \citep[e.g.,][]{M91}.   Thus, even with
their low ionization parameters, we expect the
electron temperature to be reasonably high in these HII regions.
Adopting this electron temperature and using the emission
line strengths tabulated in Table \ref{tab:LeoA}
the oxygen, nitrogen, and argon abundances
were calculated for each HII region (Table \ref{tab:abund}).
The derived mean oxygen abundance for these 4 HII regions
is 7.38 $\pm$ 0.06 $\pm$ 0.10 where the first error is
the weighted error in the mean and the second is the
systematic uncertainty due to the assumed electron
temperature. This value is in excellent agreement with both
the planetary nebula abundance (Section 3.1) and the low end of
those derived
from semi-empirical methods (Section 3.2).  Furthermore,
we emphasize that even if the derived oxygen abundances are
only rough estimates, the relative elemental abundances
are less sensitive to choice of electron temperature.

The mean log(N/O) for all 4 HII regions in
Leo~A is --1.53 $\pm$ 0.09.  
[Ar~III] was detected in only one HII region (at a
low signal--to--noise ratio), with a value of
log(Ar/O) = --2.25 $\pm$ 0.25.
Finally, since [S III] and [Ne III] were not detected,
S/O and Ne/O values were not measured for the Leo~A
HII regions.

\subsubsection{HII Regions in GR 8}

For six of the HII regions observed in GR~8, the [O~III] $\lambda$4363
line was either too weak to be detected, or was contaminated by
the nearby Hg $\lambda$4358 night sky line.  In the absence of a direct
measurement of the electron temperature, we chose to adopt a ``typical''
electron temperature of 15000 $\pm$ 2500 K to enable analysis of
nitrogen, neon, sulfur, and argon abundances.  This electron temperature
is similar to those determined in the GR~8 HII regions --019--019 and 
+008--011, and
is approximately what is necessary to reproduce the empirical
oxygen abundances of these HII regions (Table \ref{tab:abund}).
The uncertainty associated with the adopted electron temperature
includes the possibility that the spread of ionization
parameters seen in Figure \ref{fig:r23} corresponds to a spread in
electron temperatures as well.
Fortunately, the relative enrichment of nitrogen (and the other
elements) is less sensitive to the electron temperature \citep[e.g.,][]{KS96}
than a direct measurement of O/H or N/H.  The derived oxygen abundances
and log(N/O), log(Ne/O), log(S/O), and log(Ar/O) are listed in 
Table \ref{tab:abund}.

As is typical of low metallicity HII regions \citep{G90, TIL95},
the mean log(N/O) for all 8 HII regions in GR~8 is --1.51 $\pm$ 0.07.
The mean log(Ne/O) is --0.78 $\pm$ 0.17, which
is consistent with the typical values found in the high ionization
HII regions of blue compact dwarf galaxies
\citep[$<$log(Ne/O)$>$ = --0.70 $\pm$ 0.03,][]{TIL95}.
The mean log(S/O) is --1.52 $\pm$ 0.12, which again agrees favorably
with \citet{TIL95}  mean of $<$log(S/O)$>$ = --1.54 $\pm$ 0.04.
Finally, the mean log(Ar/O) is  --2.23 $\pm$ 0.08, which is
similar to the mean of $<$log(Ar/O)$>$ = --2.23 $\pm$ 0.07 found
by \citet{TIL95}.

\section{Comparison of Abundances}

\subsection{Using the PN in Leo A to Measure the ISM Abundance}

Clearly, the most secure oxygen abundance measurement in Leo~A is
that of the planetary nebula ($+$089$+$031).  However, it
is not clear that the oxygen abundance measurement in the PN is 
representative of the oxygen abundance in the present ISM.
\citet{SKH89} discussed the propriety of determining an 
ISM oxygen abundance for Leo~A from observations of its PN.
At the time, oxygen abundances measured in PN in the LMC, SMC, and 
NGC 6822 appeared to show good agreement with their corresponding
HII regions \citep{F83}.  The addition of the present Leo~A HII region 
abundance measurements appear to confirm those early results.  However,
it is possible that the agreement is coincidental.  
In this section, we examine the inferred properties of the
PN to determine whether it is justifiable to use the abundances from the 
PN to represent the ISM abundance at the present epoch. 
 We also compare the complete set of abundances of the
PN to those of the Leo~A and GR~8  HII regions to search for
systematic offsets in O, Ne, and Ar abundances.

There are two main concerns with using PN nebula abundances as an
indicator of ISM abundances.  First, if the PN is from a low mass
progenitor, and there has been significant chemical evolution
since the formation of the original star, then the PN abundances
reflect the lower abundances of a pre-enriched ISM \citep[e.g.,][]{RM95}. 
In fact, with several PN, one can establish the history of chemical
enrichment \citep[e.g.,][]{Dea97}.  Secondly,
the evolution of the progenitor star may have significantly altered the 
abundances of certain species (typically, He, C, and N).  If O is not
significantly affected by the evolution of the progenitor star,
then the PN O abundance serves as a lower limit to the ISM
abundance; further,  if a main sequence age can be estimated for the 
progenitor star, then the younger the age, the better the 
estimate of the ISM O abundance.

It is possible to estimate an age for a PN progenitor star by
making estimates of its central star T$_{eff}$ and L. 
For example, \citet{Kea05} have recently presented abundance
measurements of a PN in the nearby low metallicity galaxy
Sextans B and find good agreement between the PN O abundance 
and two of three HII regions observed. For the PN central star,
they estimate a Zanstra temperature using eqn.\ 1 of
\citet{KJ89} and a total luminosity using eqn.\ 9 of
\citet{ZP89} \citep[see also][]{GP89}.  It is then possible to use
the models of \citet{VW94} to estimate a progenitor mass and
corresponding main sequence age.  Although this requires absolute
photometry and our observations were obtained in non-photometric
conditions, the H$\alpha$ flux reported here is in good agreement 
with that of \citet{Mea03}.  Following 
this methodology for the Leo~A PN, we derive values of T$_{eff}$ = 
125,000 K and log (L/L$_\sun$) = 3.6, resulting in a progenitor mass of
$\sim$ 1 M$_\sun$ and thus a correspondingly very old age.
At face value, the agreement between the PN O abundance and those
inferred from the HII regions implies very little chemical evolution
in Leo~A over most of the lifetime of the Universe.

However, recent interest in the evolution of very low 
metallicity stars present in the early Universe has prompted
new generations of stellar evolution models at low metallicities.
One of the results of these calculations is that oxygen can
be enhanced in the envelopes of extremely metal poor AGB stars
by efficient third dredge-up \citep{H04a, H04b}.  Indeed, 
\citet{Kea05} discovered a PN in Sextans A with a 0.5 dex 
oxygen overabundance with respect to its HII regions.  With an
estimated age of $\sim$1.6 Gyr, this PN might be expected to 
have an oxygen abundance equal to or slightly lower than the 
surrounding ISM, but that expectation is clearly ruled out.  
Based on this observation,
\citet{Kea05} argue that oxygen abundances from metal poor
planetary nebula should not be used as reliable ISM abundance
indicators.  

Thus, we are forced to our alternative method of comparing Ne,
Ar, and S abundances to determine if the PN abundance
is representative of the present epoch ISM abundance.  Unfortunately, 
the lack of [OII] and [S~III]
detection in the PN and the lack of [Ne III] detections
in the Leo~A HII regions prevents a direct comparison within
the galaxy.  However, if we compare the Leo~A PN abundances
to the average of the GR~8 HII regions,
we find that the Ne abundance is lower by 0.51 $\pm$ 0.17 dex
and the Ar abundance is lower by 0.20 $\pm$ 0.14 dex.  These are
in general agreement with the average offset in the semi-empirical
oxygen abundances of 0.29 dex.  

In sum, it would appear that the PN in Leo~A probably has a low
mass progenitor, and therefore is less than ideal for making an
estimate of the present day ISM oxygen abundance.  It is also
clearly a very low metallicity PN, and thus, subject to oxygen
enrichment through stellar evolutionary processes.  Nonetheless,
the abundances derived from the PN agree with those of
the HII regions, and has the benefit of having [O~III] $\lambda$4363
detected.  While a higher s/n spectrum which detects the lower
ionization species and the near-IR [S~III] lines remains
highly desirable, we adopt the PN abundance as representative
of the ISM abundance in Leo~A for the remainder of this paper.

\subsection{Oxygen in Leo A and GR 8: Global Abundance Values 
and Chemical Mixing}

Both Leo~A (12 + log(O/H) = 7.30) and GR~8 (12 + log(O/H) = 7.65)
have extremely low oxygen abundances,\footnote{Leo A's global oxygen
abundance is derived from observations of a planetary nebula;
GR~8's global oxygen abundance is derived from the weighted average
of the two HII regions with direct abundance measurements.}
 on the order of 3--5\% of 
the solar value \citep[using the old oxygen scale,][]{L78} or
5--10\% \citep[using the new scale,][]{APLA01, Aea04, M04}.
Similar extremely low oxygen abundances are seen in other low
luminosity galaxies, such as 
UGCA 292 \citep[12 + log(O/H) = 7.30,][]{vZ00},
SagDIG \citep[12 + log(O/H) = 7.44,][]{SKH89, Se02}, 
and UGC 4483 \citep[12 + log(O/H) = 7.56,][]{STKGT94,vH04}; 
however, all are more metal--rich
than I Zw 18 \citep[12 + log(O/H) = 7.20,][]{SS72,SK93}.

At these very low abundances, and given the lack of shear which
would lead to more efficient mixing, it would be natural to 
expect significant chemical inhomogeneities in the ISM of 
these dwarf galaxies \citep{SKH89}.  Indeed, one of the 
motivations for this study was the anomalously high N/O ratio 
observed in one of the HII regions in GR~8 by \citet{MAM90}.  
However, we find no evidence of N/O enrichment in
any of the HII regions observed.  Further, the derived oxygen 
abundances appear to be uniform in both galaxies 
(within the measurement errors), indicating that
the enriched material is well mixed in both of these 
low mass systems.  In fact, within the accuracy of the
measurements, we find no evidence of abundance variations 
in any of the elements observed in
either of the two galaxies.  Perusal of Tables 5 and 6 indicates
that excursions from the average O/H and N/O of more than 0.1 
and 0.15 dex (respectively) seem unlikely.  Within errors, all of
the abundance measurements appear to be consistent with 
uniform abundances.

In the past, there have been insufficient HII region measurements to
provide a robust test for abundance dispersions in low luminosity dwarf
irregular galaxies.  However, with the availability of spectrographs on
larger telescopes, the situation is improving 
\citep[e.g.,][]{M96,KS96,KS97,VP98,LS04,vH04,LSV05a,LSV05b}.
Additionally, abundances for individual young stars are now becoming
available \citep[e.g.,][]{Vea01, Vea03, Kea04}.  
Interestingly, based on observations of only 3 HII regions
in Sextans B, \citet{Kea05} have identified an oxygen abundance difference
of $\sim$ 0.3 dex.
Nonetheless, at present, it appears that significant departures 
from a uniform ISM abundance are the exception and not the 
rule \citep[e.g., NGC 5253,][]{Kea97}.  The lack of significant
dispersion in abundance even at these extremely low metallicities
supports the picture of efficient mixing of newly synthesized elements into 
the hot phase of the ISM before cooling back down to the observable 
warm phase of the ISM \citep[e.g.,][]{T96, KS97, Lea00}.

\subsection{The Metallicity -- Luminosity Relationship}

Leo~A and GR~8 contribute significantly to the number
of low luminosity, low metallicity galaxies known.
Using the compilation of \citet{KKHM04} to identify
galaxies within the local 5 Mpc volume, we have compiled a list
of all nearby low luminosity galaxies with known oxygen abundances.
The 5 Mpc volume was chosen to include several of the nearest
dwarf rich groups (e.g., Local Group, M81, Sculptor, and Centaurus).
Of the 163 gas-rich galaxies (morphological type $>$ 0) in this volume,
144 have M$_{\rm B} > -18$.  Of these, only 50 have measured oxygen
abundances in the literature (Table \ref{tab:lit}).  Also included
in Table \ref{tab:lit} are the optical magnitudes and distances
compiled by \citet{KKHM04} and an indication of the methods used
to calculate the distance \citep{KKHM04} and the oxygen and nitrogen abundances.
The luminosity--metallicity relationship for the local volume 
is shown in Figure \ref{fig:metlum}.
In creating Figure \ref{fig:metlum}, we have excluded PegDIG
because the oxygen abundance is based on measurements of only [O~II] and
thus may not be consistent with the other abundance calibrations.

A weighted least squares fit to the data yields:
\begin{equation}
12 + {\rm log(O/H)} = 5.67 - 0.151 {\rm M_B}
\end{equation}
with an error in the slope and zeropoint of 0.014 and 0.21,
respectively.
This result is remarkably similar to that presented by \citet{SKH89}; 
the slight difference in the zero point (5.67 instead of 5.60) can 
easily be attributed to the slightly different distance scales 
used for the compilations.
Similar results are also found by \citet{RM95}, 
\citet{vHS97b}, \citet{Le03a}, and \citet{vH04}.

Figure \ref{fig:metlum} gives the impression that the dispersion is
larger for the empirically derived oxygen abundances relative to those
derived via the direct method.  Statistically, this is correct.
For the 33 galaxies with direct oxygen abundances, the dispersion
about the relationship is $\sigma$ $=$ 0.17 dex, and for the 
16 galaxies with empirical oxygen abundances, the dispersion is
$\sigma$ $=$ 0.29 dex.  However, this could be misleading.
Three of the galaxies in the compilation have empirical oxygen abundances
based on relatively lower quality spectra (UGCA 86, DDO 168, NGC 5264).
(Note that when [O III] $\lambda$4363 is measured in a spectrum, it is, by
definition, a higher quality spectrum.)
When these three galaxies are removed from the empirical oxygen abundance
sample, the dispersion decreases to $\sigma$ $=$ 0.19 dex -- nearly
identical to that of the direct abundance sample.
This implies that the intrinsic scatter in the metallicity--luminosity
relationship is at least of order the size of the uncertainty
in the abundance measurements.  Given their apparent uncertanties, 
we delete these three galaxies from subsequent analysis of
the metallicity-luminosity relationship.

\citet{RM95} found an increased scatter in the metallicity--luminosity
relationship at lower luminosities, with an abrupt onset at M$_B$ $=$ $-$15.
If we divide our data into a high luminosity sample (M$_B$ $\le$ $-$15)
and a low luminosity sample, for the 21 high luminosity galaxies we
calculate a dispersion of $\sigma$ $=$ 0.19 dex and for the 25 low luminosity
galaxies a dispersion of $\sigma$ $=$ 0.16 dex.  Thus, we find no
evidence for an increased dispersion at lower luminosities \citep[see also][]{Le03a}.  
This is surprising, as one expects that the lower luminosity galaxies should
be susceptible to larger departures in both abundance and luminosity.
Note that Leo~A and GR~8 play a significant role in the differences
between the two studies.  Although the abundances for these two galaxies
are nearly identical in the two studies, the distance to GR~8 is now
known to be roughly double and the distance to Leo~A is roughly half 
those used by \citet{RM95}.

\section{Low O/H as a Young Galaxy Hypothesis}

Based on spectroscopic observations of HII regions in low-metallicity
blue compact galaxies, \citet{IT99} hypothesized that ``galaxies with
12 $+$ log (O/H) $\le$ 7.6 are now undergoing their first burst of
star formation, and that they are therefore young, with ages not
exceeding 40 Myr.''  This hypothesis is based on observations
of nearly constant N/O and C/O ratios for their sample of galaxies.
The reasoning holds that these galaxies have not had sufficient time
for the intermediate mass stars to deliver their time-delayed 
production of N and C.
In Leo~A, we have excellent evidence that 12 $+$ log (O/H) $\le$ 7.6
and the value of log(N/O) $=$ $-$1.53 $\pm$ 0.09 is consistent with
their remarkably narrow plateau of $-$1.60 $\pm$ 0.02. 
Leo~A thus provides an excellent test case for the young
galaxy hypothesis.

There is abundant evidence that Leo~A is not a young
galaxy.  The Hubble Space Telescope color-magnitude diagram presented
by \citet{Te98} shows very well populated red giant branch and
red clumps, indicative of intermediate and old age stars.
Further, \citet{Dea02} discovered 8 RR Lyrae stars indicative of a $\sim$ 10
Gyr old stellar population.  Thus, although Leo~A has not necessarily been 
forming stars at a constant rate over the lifetime of the Universe 
\citep{Te98}, it has clearly has stars with a wide variety of ages.

\citet{Mea05} come to a very similar conclusion in their
study of Sag DIG.   In fact, an ancient stellar population has been 
detected in all low metallicity dwarf galaxies with sufficiently 
deep observations of their resolved stellar populations \citep{M98}.
The one possible exception has been I~Zw~18;  \citet{IT04} claim an 
absence of any old stellar population based on new Hubble Space Telescope imaging. 
However, \citet{Mea05} have reanalyzed
the same Hubble Space Telescope observations, and, based on photometry
which reaches roughly one magnitude deeper than that of \citet{IT04},
find evidence consistent with the detection of a red giant branch
tip corresponding to a distance of 15 Mpc.  Although not noted, 
the extended red supergiant branch is also not consistent with 
an age of less than 40 Myr for all of the resolved stars.  Thus, it would 
appear that, to date, there is no evidence in support of dwarf galaxies being 
formed for the first time in the current epoch.

\section{Conclusions}
We present the results of optical spectroscopy of 4 HII regions and
one planetary nebula in Leo~A and 8 HII regions in GR~8.
Observations of the planetary nebula in Leo~A yield 12 + log(O/H) = 
7.30 $\pm$ 0.05 in agreement with ``semi-empirical'' calculations
of the oxygen abundance in its HII regions yielding
12 + log(O/H) = 7.38 $\pm$ 0.10.
These results confirm that Leo~A has one of the lowest ISM metal abundances
of known nearby galaxies.
From two HII regions with [O~III] $\lambda$4363 detections, 
the mean oxygen abundance of GR~8 is 12 + log(O/H) = 7.65 $\pm$ 0.06,
in agreement with ``empirical'' and
``semi-empirical'' abundances for the 6 other HII regions.
Similar to previous results in other low metallicity galaxies,
the mean log(N/O) $=$ $-$1.53 $\pm$ 0.09 for Leo~A and
$-$1.51 $\pm$ 0.07 for GR~8.
There is no evidence of significant variations in either O/H or
N/O in the HII regions.

The metallicity-luminosity relation for nearby (D $<$ 5 Mpc) dwarf
irregular galaxies with measured oxygen abundances has a mean correlation
of 12 + log(O/H) = 5.67 $-$ 0.151 M$_{\rm B}$ with a dispersion in
oxygen about the relationship of $\sigma$ $=$ 0.21.
These observations confirm that gas-rich low luminosity galaxies have
extremely low elemental abundances in the ionized gas-phase of their
interstellar media.

Although Leo~A has one of the lowest metal abundances of known nearby
galaxies, detection of tracers of an older stellar population (RR Lyrae
variable stars, horizontal branch stars, a well populated red giant
branch) indicate that it is not a newly formed galaxy as has been
proposed for some other similarly low metallicity star forming galaxies
(e.g., I~Zw~18, SBS 0335$-$052).  Because Leo~A has very similar ISM
abundances to these systems, it could be taken as evidence against the
hypothesis that these are young galaxies.

\acknowledgments
We thank Eline Tolstoy and John Salzer for providing optical and H$\alpha$
images of Leo~A and GR~8.  We also thank Chip Kobulnicky for assistance
in selecting the HII region targets in Leo~A.  We thank Henry Lee for
several valuable comments on this manuscript.
This research has made use of the NASA/IPAC Extragalactic Database (NED)
which is operated by the Jet Propulsion Laboratory, California Institute
of Technology, under contract with the National Aeronautics and Space
Administration.  LvZ acknowledges partial support from Indiana University.
EDS is grateful for partial support from a NASA LTSARP grant No. NAG5-9221
and the University of Minnesota. MPH has been supported by NSF grants AST-9900695
and AST-0307396.

\begin{deluxetable}{llcccccccc}
\tablewidth{0pt}
\tabcolsep 3pt
\tablecaption{Global Parameters\label{tab:global}}
\tablehead{
\colhead{} &\colhead{RA}&\colhead{Dec}&\colhead{D}& \colhead{d$_{25}$}& \colhead{} & \colhead{}  & \colhead{\underline{M$_{\rm HI}$}}& \colhead{12 +} \\[.2ex]
\colhead{Galaxy}&\colhead{(2000)}& \colhead{(2000)} &\colhead{[Mpc]}& \colhead{[kpc]}& \colhead{M$_{\rm B}$} & \colhead{(B-V)$_0$}  & \colhead{L$_{\rm B}$ }& \colhead{log(O/H)} & \colhead{log(N/O)}}
\startdata
Leo A  & 09 59 24.8 & 30 44 57 & 0.69 & 1.0  & --11.52 & 0.26 & 1.3 & 7.30 $\pm$ 0.05  & --1.53 $\pm$ 0.09\\
GR 8   & 12 58 39.8 & 14 13 07 & 2.20 & 0.69 & --12.12 & 0.38 & 0.76 & 7.65 $\pm$ 0.06 & --1.51 $\pm$ 0.07\\
\enddata
\tablecomments{All parameters except 12+log(O/H) and log(N/O) from Dohm--Palmer et al.\ 1998 and references therein. 12 + log(O/H) values are from direct abundance 
measurements (Table 5) and log (N/O) values are from semi-empirical measurements
(Table 6).}
\end{deluxetable}

\begin{deluxetable}{lrrrr}
\tablewidth{0pt}
\tablecaption{Observing Log\label{tab:obs}}
\tablehead{
\colhead{Slit } & \colhead{RA} &\colhead{Dec}& \colhead{PA} & \colhead{T$_{\rm int}$} \\
\colhead{Position}& \colhead{(2000)} & \colhead{(2000)} & \colhead{[deg]}& \colhead{[sec]}
}
\startdata
Leo A/A & 09 59 31.8 & 30 44 37 & 92 & 7 $\times$ 1200 \\ 
Leo A/B & 09 59 31.6 & 30 45 28 & 90 & 3 $\times$ 1200 \\ 
Leo A/C & 09 59 17.2 & 30 44 07 & 73 & 6 $\times$ 1200 \\ 
GR 8/A  & 12 58 38.5 & 14 12 49 &  0 & 3 $\times$ 1200 \\ 
GR 8/B  & 12 58 40.1 & 14 13 01 & 45 & 2 $\times$ 1200 \\ 
GR 8/C  & 12 58 39.9 & 14 13 34 &  0 & 4 $\times$ 1200 \\ 
GR 8/D  & 12 58 40.4 & 14 12 56 & 45 & 2 $\times$ 1200 \\ 
\enddata
\end{deluxetable}

\begin{deluxetable}{lcccccc}
\tablewidth{0pt}
\tabletypesize{\footnotesize}
\tabcolsep 3pt
\tablecaption{Optical Line Intensities for HII Regions in Leo A
\label{tab:LeoA}}
\tablehead{
\colhead{Ionic} & \colhead{Rest} & \colhead{--101--052}& \colhead{--091--048} & \colhead{+069--018} & \colhead{+089+031 (PN)}& \colhead{+112--020} \\
\colhead{Species} &\colhead{Wavelength} &\colhead{I($\lambda$)/I(H$\beta$)}&\colhead{I($\lambda$)/I(H$\beta$)}  &\colhead{I($\lambda$)/I(H$\beta$)} &\colhead{I($\lambda$)/I(H$\beta$)} &\colhead{I($\lambda$)/I(H$\beta$)} 
}
\startdata
{[OII]}   &  3727 & 1.260$\pm$0.077 & 1.288$\pm$0.086 & 1.834$\pm$0.131 & $<$ 0.080     & 1.767$\pm$0.080 \\
{[NeIII]} &  3869 &      \nodata    &      \nodata    &      \nodata    & 0.266$\pm$0.019 &    \nodata      \\
H$\gamma$ &  4340 & 0.474$\pm$0.036 & 0.419$\pm$0.039 & 0.432$\pm$0.048 & 0.441$\pm$0.021 & 0.425$\pm$0.022 \\
{[OIII]}  &  4363 &      \nodata    &      \nodata    &      \nodata    & 0.137$\pm$0.014 &    \nodata      \\
HeI       &  4471 &      \nodata    &      \nodata    &      \nodata    & 0.046$\pm$0.013 &    \nodata      \\
HeII      &  4686 &      \nodata    &      \nodata    &      \nodata    & 0.212$\pm$0.015 &    \nodata      \\
H$\beta$  &  4861 & 1.000$\pm$0.050 & 1.000$\pm$0.056 & 1.000$\pm$0.066 & 1.000$\pm$0.035 & 1.000$\pm$0.036 \\
{[OIII]}  &  4959 & 0.443$\pm$0.034 &      \nodata    &      \nodata    & 1.279$\pm$0.044 &    \nodata      \\
{[OIII]}  &  5007 & 1.055$\pm$0.052 & 0.672$\pm$0.045 & 0.105$\pm$0.042 & 3.781$\pm$0.124 & 0.296$\pm$0.018 \\
H$\alpha$ &  6563 & 2.743$\pm$0.161 & 2.225$\pm$0.142 & 2.775$\pm$0.202 & 2.701$\pm$0.125 & 2.537$\pm$0.120 \\
{[NII]}   &  6584 & 0.035$\pm$0.007 & 0.058$\pm$0.010 & 0.084$\pm$0.012 & 0.023$\pm$0.003 & 0.061$\pm$0.005 \\
HeI       &  6678 & 0.042$\pm$0.007 &      \nodata    &      \nodata    & 0.040$\pm$0.004 &    \nodata      \\
{[SII]}   &  6716 & 0.114$\pm$0.010 & 0.107$\pm$0.012 & 0.172$\pm$0.017 &      \nodata    & 0.140$\pm$0.008 \\
{[SII]}   &  6731 & 0.069$\pm$0.008 & 0.070$\pm$0.010 & 0.119$\pm$0.014 &      \nodata    & 0.101$\pm$0.006 \\
HeI       &  7065 &      \nodata    &      \nodata    &      \nodata    & 0.125$\pm$0.007 & 0.027$\pm$0.004 \\
{[ArIII]} &  7136 &      \nodata    &      \nodata    &      \nodata    & 0.011$\pm$0.003 & 0.008$\pm$0.004 \\
\\
\hline
\\
\multicolumn{2}{l}{c$_{H\beta}$}                  &  0.00$\pm$ 0.06& 0.00$\pm$ 0.06& 0.00$\pm$ 0.07& 0.16$\pm$ 0.05& 0.00$\pm$ 0.05\\
\multicolumn{2}{l}{F(H$\beta$)$\times10^{15}$}    &   0.76 &   0.52  &   0.70  &   1.88  &   1.24 \\
\multicolumn{2}{l}{EW(H$\beta$) [\AA]}            &  280& 306&  13&1607& 103\\
\multicolumn{2}{l}{Log(([OII]+[OIII])/H$\beta$)}  & 0.426$\pm$0.016&  0.339$\pm$0.020&  0.295$\pm$0.030&  0.704$\pm$0.011&  0.335$\pm$0.017\\
\multicolumn{2}{l}{Log([OIII]/[OII])}            & 0.048$\pm$0.033& -0.158$\pm$0.040& -1.117$\pm$0.138&       \nodata   & -0.651$\pm$0.032\\
Slit      &       &      C          &          C      &         A       &         B       &        A        
\enddata
\end{deluxetable}

\begin{deluxetable}{lccccccccc}
\tabletypesize{\scriptsize}
\rotate
\tablewidth{0pt}
\tabcolsep 2pt
\tablecaption{Optical Line Intensities for HII Regions in GR 8
\label{tab:GR8}}
\tablehead{
\colhead{Ionic} & \colhead{Rest} & \colhead{--019--019}& \colhead{--013--032} & \colhead{--012--022} & \colhead{--002--012}& \colhead{+001--008} & \colhead{+001+027} & \colhead{+004--006} & \colhead{+008--011} \\
\colhead{Species} &\colhead{Wavelength} &\colhead{I($\lambda$)/I(H$\beta$)}&\colhead{I($\lambda$)/I(H$\beta$)}  &\colhead{I($\lambda$)/I(H$\beta$)} &\colhead{I($\lambda$)/I(H$\beta$)} &\colhead{I($\lambda$)/I(H$\beta$)} &\colhead{I($\lambda$)/I(H$\beta$)} &\colhead{I($\lambda$)/I(H$\beta$)} &\colhead{I($\lambda$)/I(H$\beta$)} 
}
\startdata
{[OII]}   &  3727 & 2.031$\pm$0.084 & 2.581$\pm$0.170 & 1.536$\pm$0.068 & 2.305$\pm$0.112 & 2.602$\pm$0.109 & 2.362$\pm$0.134 & 2.642$\pm$0.106 & 2.160$\pm$0.089 \\
{[NeIII]} &  3869 & 0.157$\pm$0.011 &     \nodata     & 0.228$\pm$0.016 &     \nodata     & 0.169$\pm$0.014 &    \nodata      & 0.093$\pm$0.008 & 0.245$\pm$0.013 \\
H$\gamma$ &  4340 & 0.501$\pm$0.019 & 0.467$\pm$0.046 & 0.359$\pm$0.018 & 0.381$\pm$0.023 & 0.471$\pm$0.020 & 0.566$\pm$0.039 & 0.421$\pm$0.016 & 0.427$\pm$0.017 \\
{[OIII]}  &  4363 & 0.045$\pm$0.008 &     \nodata     &     \nodata     &     \nodata     &     \nodata     &    \nodata      &     \nodata     & 0.038$\pm$0.008 \\
HeI       &  4471 & 0.023$\pm$0.008 &     \nodata     &     \nodata     &     \nodata     &     \nodata     &    \nodata      & 0.037$\pm$0.007 &     \nodata     \\
H$\beta$  &  4861 & 1.000$\pm$0.032 & 1.000$\pm$0.061 & 1.000$\pm$0.034 & 1.000$\pm$0.040 & 1.000$\pm$0.033 & 1.000$\pm$0.049 & 1.000$\pm$0.031 & 1.000$\pm$0.032 \\
{[OIII]}  &  4959 & 0.675$\pm$0.022 & 0.613$\pm$0.047 & 0.970$\pm$0.034 & 0.419$\pm$0.023 & 0.449$\pm$0.018 & 0.537$\pm$0.035 & 0.387$\pm$0.014 & 0.619$\pm$0.021 \\
{[OIII]}  &  5007 & 2.103$\pm$0.065 & 1.552$\pm$0.084 & 2.951$\pm$0.095 & 1.194$\pm$0.046 & 1.444$\pm$0.046 & 1.623$\pm$0.072 & 1.131$\pm$0.035 & 1.934$\pm$0.059 \\
{[SIII]}  &  6311 & 0.016$\pm$0.003 &     \nodata     & 0.017$\pm$0.003 & 0.023$\pm$0.004 &     \nodata     &    \nodata      & 0.013$\pm$0.002 & 0.014$\pm$0.003 \\
H$\alpha$ &  6563 & 2.590$\pm$0.113 & 2.786$\pm$0.190 & 2.586$\pm$0.118 & 2.502$\pm$0.125 & 2.803$\pm$0.125 & 2.872$\pm$0.166 & 2.563$\pm$0.110 & 2.712$\pm$0.118 \\
{[NII]}   &  6584 & 0.078$\pm$0.005 & 0.139$\pm$0.013 & 0.076$\pm$0.004 & 0.115$\pm$0.007 & 0.084$\pm$0.006 & 0.081$\pm$0.008 & 0.109$\pm$0.005 & 0.094$\pm$0.005 \\
HeI       &  6678 & 0.021$\pm$0.003 &     \nodata     & 0.019$\pm$0.003 & 0.029$\pm$0.004 & 0.026$\pm$0.005 &    \nodata      & 0.021$\pm$0.002 & 0.017$\pm$0.003 \\
{[SII]}   &  6716 & 0.153$\pm$0.008 & 0.298$\pm$0.023 & 0.163$\pm$0.008 & 0.274$\pm$0.015 & 0.255$\pm$0.013 & 0.173$\pm$0.013 & 0.231$\pm$0.011 & 0.165$\pm$0.008 \\
{[SII]}   &  6731 & 0.102$\pm$0.006 & 0.181$\pm$0.016 & 0.113$\pm$0.006 & 0.183$\pm$0.010 & 0.165$\pm$0.009 & 0.110$\pm$0.010 & 0.163$\pm$0.008 & 0.114$\pm$0.006 \\
HeI       &  7065 & 0.021$\pm$0.003 &     \nodata     & 0.015$\pm$0.003 &     \nodata     &     \nodata     &    \nodata      & 0.017$\pm$0.002 & 0.026$\pm$0.003 \\
{[ArIII]} &  7136 & 0.042$\pm$0.004 & 0.043$\pm$0.009 & 0.048$\pm$0.004 &     \nodata     & 0.045$\pm$0.005 & 0.037$\pm$0.007 & 0.042$\pm$0.003 & 0.056$\pm$0.004 \\
\\
\hline
\\
\multicolumn{2}{l}{c$_{H\beta}$}                  &  0.00$\pm$ 0.05& 0.12$\pm$ 0.07& 0.00$\pm$ 0.05& 0.00$\pm$ 0.06 & 0.15$\pm$ 0.05& 0.22$\pm$ 0.06&  0.00$\pm$ 0.05& 0.06$\pm$ 0.04 \\
\multicolumn{2}{l}{F(H$\beta$)$\times10^{15}$}    &   2.87 &  1.57 &  4.95 &  2.12 &  3.73 &  0.78 &  7.46 &  5.98 \\
\multicolumn{2}{l}{EW(H$\beta$) [\AA]}            &   60&  40&  58&  66&  63&  30&  63&  27 \\
\multicolumn{2}{l}{Log(([OII]+[OIII])/H$\beta$)}  &  0.682$\pm$0.010& 0.676$\pm$0.018& 0.737$\pm$0.010& 0.593$\pm$0.014& 0.653$\pm$0.012& 0.655$\pm$0.015& 0.619$\pm$0.012& 0.673$\pm$0.010\\
\multicolumn{2}{l}{Log([OIII]/[OII])}            &   0.136$\pm$0.021& -0.076$\pm$0.035&  0.407$\pm$0.022& -0.155$\pm$0.025& -0.138$\pm$0.021& -0.039$\pm$0.029& -0.241$\pm$0.021&  0.073$\pm$0.021\\
Slit      &       &      A          &          D      &         B       &         B       &        C       &     C     & B    &  D    
\enddata
\end{deluxetable}

\begin{deluxetable}{lrcccc}
\tablewidth{0pt}
\tabletypesize{\footnotesize}
\tabcolsep 3pt
\tablecaption{Derived Abundances for Bright HII Regions in Leo A and GR 8
\label{tab:new}}
\tablehead{
\colhead{Species}  & & \colhead{Leo A +089+031 (PN)} & \colhead{GR 8 --019--019} &\colhead{GR 8 +008--011} }
\startdata
(N$^+$/H)     & & 1.63 $\pm$ 0.28 $\times10^{-7}$ & 5.82 $\pm$ 0.70 $\times10^{-7}$ & 8.67 $\pm$ 0.95 $\times10^{-7}$ \\
          & ICF &  \nodata                        & 1.91 $\pm$ 0.39                 & 1.82 $\pm$ 0.40                 \\
(N/H)         & &  \nodata                        & 1.11 $\pm$ 0.26 $\times10^{-6}$ & 1.57 $\pm$ 0.39 $\times10^{-6}$ \\
(O$^+$/H)     & &  $<$ 8.7 $\times10^{-7}$          & 2.21 $\pm$ 0.37 $\times10^{-5}$ & 2.49 $\pm$ 0.45 $\times10^{-5}$ \\
(O$^{++}$/H)  & & 1.98 $\pm$ 0.22 $\times10^{-5}$ & 2.02 $\pm$ 0.32 $\times10^{-5}$ & 2.04 $\pm$ 0.36 $\times10^{-5}$ \\
(O/H)         & & 1.98 $\pm$ 0.22 $\times10^{-5}$ & 4.23 $\pm$ 0.49 $\times10^{-5}$ & 4.53 $\pm$ 0.58 $\times10^{-5}$ \\
(Ne$^{++}$/H) & & 2.37 $\pm$ 0.45 $\times10^{-6}$ & 2.80 $\pm$ 0.76 $\times10^{-6}$ & 4.87 $\pm$ 1.45 $\times10^{-6}$ \\
          & ICF & 1.00 $\pm$ 0.16                 & 2.10 $\pm$ 0.41                 & 2.23 $\pm$ 0.48                 \\
(Ne/H)        & & 2.37 $\pm$ 0.58 $\times10^{-6}$ & 5.87 $\pm$ 1.95 $\times10^{-6}$ & 1.08 $\pm$ 0.40 $\times10^{-5}$ \\
(S$^+$/H)     & &  \nodata                        & 2.84 $\pm$ 0.27 $\times10^{-7}$ & 3.23 $\pm$ 0.32 $\times10^{-7}$ \\
(S$^{++}$/H)  & &  \nodata                        & 7.36 $\pm$ 2.47 $\times10^{-7}$ & 7.24 $\pm$ 2.73 $\times10^{-7}$ \\ 
          & ICF &  \nodata                        & 1.19 $\pm$ 0.12                 & 1.18 $\pm$ 0.12                 \\
(S/H)         & &  \nodata                        & 1.21 $\pm$ 0.32 $\times10^{-6}$ & 1.23 $\pm$ 0.35 $\times10^{-6}$ \\
(Ar$^{++}$/H) & & 0.26 $\pm$ 0.08 $\times10^{-7}$ & 1.56 $\pm$ 0.32 $\times10^{-7}$ & 2.22 $\pm$ 0.47 $\times10^{-7}$ \\
          & ICF & \nodata                         & 1.48 $\pm$ 0.15                 & 1.50 $\pm$ 0.15                 \\
(Ar/H)        & & \nodata                         & 2.30 $\pm$ 0.52 $\times10^{-7}$ & 3.34 $\pm$ 0.78 $\times10^{-7}$ \\
\\
\hline
\\
T(O$^{++}$)    & & 20960 $\pm ^{1600}_{1280}$ & 15790 $\pm ^{1690}_{1170}$   &  15190 $\pm ^{1860}_{1220}$ \\
T(O$^{+}$)     & & 15660 $\pm$ 430            & 13950 $\pm$ 600              &  13720 $\pm$ 700  \\
N$_e$          & & $\le 100$ cm$^{-3}$        & $\le 100$ cm$^{-3}$          &  $\le 100$ cm$^{-3}$ \\
log(N/H)  + 12 & &   \nodata                  & 6.05 $\pm$ 0.10              & 6.20 $\pm$ 0.11 \\
log(O/H)  + 12 & & 7.30 $\pm$ 0.05            & 7.63 $\pm$ 0.05              & 7.66 $\pm$ 0.06 \\
log(Ne/H) + 12 & & 6.38 $\pm$ 0.11            & 6.77 $\pm$ 0.14              & 7.03 $\pm$ 0.16 \\
log(S/H)  + 12 & &   \nodata                  & 6.08 $\pm$ 0.11              & 6.09 $\pm$ 0.12 \\
log(Ar/H) + 12 & &   \nodata                  & 5.36 $\pm$ 0.10              & 5.52 $\pm$ 0.10 \\
log(N/O)       & &   \nodata                  & -1.58 $\pm$ 0.11              & -1.46 $\pm$ 0.12 \\
log(Ne/O)      & & -0.92 $\pm$ 0.12           & -0.86 $\pm$ 0.15              & -0.63 $\pm$ 0.17 \\
log(S/O)       & &   \nodata                  & -1.55 $\pm$ 0.12              & -1.57 $\pm$ 0.13 \\
log(Ar/O)      & &   \nodata                  & -2.27 $\pm$ 0.11              & -2.14 $\pm$ 0.11 
\enddata
\end{deluxetable}

\begin{deluxetable}{lrccccccc}
\tabletypesize{\footnotesize}
\tablewidth{0pt}
\tabcolsep 3pt
\tablecaption{HII Region Semi-Empirical Abundances\label{tab:abund}}
\tablehead{
\colhead{Galaxy} & \colhead{Offsets} & \colhead{12+Log(O/H)}& \colhead{12+Log(O/H)}& \colhead{Log(N/O)} & \colhead{Log(Ne/O)} & \colhead{Log(S/O)} & \colhead{Log(Ar/O)} \\
 \colhead{}& \colhead{E~~~N } & \colhead{Empirical} & \colhead{Semi-Empirical}
}
\startdata
Leo A   &--101--052 &  7.48&  7.44$\pm$ 0.10& -1.66$\pm$ 0.18&     \nodata    & \nodata        &  \nodata \\
Leo A   &--091--048 &  7.45&  7.36$\pm$ 0.11& -1.45$\pm$ 0.17&     \nodata    & \nodata        &  \nodata \\
Leo A   & +069--018 &  7.70&  7.36$\pm$ 0.10& -1.44$\pm$ 0.12&     \nodata    & \nodata        &  \nodata \\
Leo A   & +112--020 &  7.65&  7.38$\pm$ 0.11& -1.56$\pm$ 0.16&     \nodata    & \nodata        & -2.25$\pm$ 0.25 \\
$<$Leo A$>$  & \nodata &  7.58&  7.38$\pm$ 0.06& -1.53$\pm$ 0.09&     \nodata    & \nodata        & -2.25$\pm$ 0.25 \\
GR 8    &--019--019 &  7.78&  7.67$\pm$ 0.10& -1.61$\pm$ 0.15& -0.85$\pm$ 0.28& -1.53$\pm$ 0.23& -2.27$\pm$ 0.18 \\
GR 8    &--013--032 &  7.85&  7.69$\pm$ 0.10& -1.37$\pm$ 0.16&     \nodata    &    \nodata     & -2.24$\pm$ 0.20 \\
GR 8    &--012--022 &  7.77&  7.70$\pm$ 0.10& -1.41$\pm$ 0.16& -0.83$\pm$ 0.28& -1.50$\pm$ 0.23& -2.24$\pm$ 0.19 \\
GR 8    &--002--012 &  7.75&  7.64$\pm$ 0.10& -1.40$\pm$ 0.16&     \nodata    & -1.34$\pm$ 0.22&  \nodata \\
GR 8    & +001--008 &  7.83&  7.70$\pm$ 0.10& -1.59$\pm$ 0.16& -0.64$\pm$ 0.29&   \nodata      & -2.24$\pm$ 0.18 \\
GR 8    &  +001+027 &  7.80&  7.66$\pm$ 0.10& -1.57$\pm$ 0.16&     \nodata    &   \nodata      & -2.28$\pm$ 0.20 \\
GR 8    & +004--006 &  7.83&  7.66$\pm$ 0.10& -1.49$\pm$ 0.14& -0.82$\pm$ 0.29& -1.55$\pm$ 0.20& -2.17$\pm$ 0.18 \\
GR 8    & +008--011 &  7.78&  7.67$\pm$ 0.10& -1.46$\pm$ 0.14& -0.62$\pm$ 0.28& -1.56$\pm$ 0.22& -2.14$\pm$ 0.18 \\
$<$GR 8$>$  & \nodata &  7.80&  7.67$\pm$ 0.04& -1.51$\pm$ 0.07& -0.78$\pm$ 0.17& -1.52$\pm$ 0.12& -2.23$\pm$ 0.08 
\enddata
\tablecomments{Empirical oxygen abundances are derived from the strong-line diagnostics of McGaugh (1991)
and have typical uncertainties of 0.1-0.2 dex.
The abundance ratios and semi-empirical oxygen abundances are calculated by adopting T$_{\rm [OIII]}$ = 15000 $\pm$ 2500 K
as a typical electron temperature for low metallicity HII regions.}
\end{deluxetable}

\begin{deluxetable}{lrrcccccc}
\tabletypesize{\tiny}
\tablewidth{0pt}
\tablecaption{Oxygen and Nitrogen Abundances for Low Luminosity Galaxies (D $<$ 5 Mpc)
\label{tab:lit}}
\tablehead{
\colhead{Galaxy} & \colhead{m$_{\rm B}$} & \colhead{Distance}&\colhead{D$_{\rm method}$}& \colhead{M$_{\rm B}$}& \colhead{12+Log(O/H)} & \colhead{Log(N/O)}& \colhead{(O/H)$_{\rm method}$} & \colhead{Reference} \\
\colhead{} & \colhead{} & \colhead{[Mpc]} \\
}
\startdata
       Leo A& 12.92 & 0.69  & rgb & -11.36 & 7.30 $\pm$ 0.05 & -1.53 $\pm$ 0.09 & Direct (PN)& 0,1  \\
    UGCA 292& 16.10 & 3.1   & bs  & -11.43 & 7.30 $\pm$ 0.03 & -1.45 $\pm$ 0.07 & Direct    &  2,3  \\
     Pegasus& 13.21 & 0.76  & rgb & -11.47 & 7.93 $\pm$ 0.14 & -1.24 $\pm$ 0.15 & [OII]     &  4    \\
     Sag DIG& 14.12 & 1.04  & rgb & -11.49 & 7.44 $\pm$ 0.20 & -1.63 $\pm$ 0.20 & Empirical & 5,6,7 \\
        GR 8& 14.61 & 2.1   & rgb & -12.11 & 7.65 $\pm$ 0.06 & -1.51 $\pm$ 0.07 & Direct    & 0,8,9,10\\
     HIZSS 3& 18.00 & 1.4   & h   & -12.39 & 7.80 $\pm$ 0.20 &      \nodata     & Empirical &  11   \\
     DDO 167& 15.45 & 4.19  & rgb & -12.70 & 7.60 $\pm$ 0.20 &      \nodata     & Empirical &  1,10 \\
    UGC 9128& 14.38 & 2.5   & rgb & -12.71 & 7.75 $\pm$ 0.05 & -1.80 $\pm$ 0.12 & Direct    & 1,12,13\\
    UGC 4483& 14.95 & 3.21  & rgb & -12.73 & 7.56 $\pm$ 0.03 & -1.57 $\pm$ 0.07 & Direct    &  3,14  \\
     DDO 181& 14.45 & 3.01  & rgb & -12.97 & 7.85 $\pm$ 0.04 & -1.60 $\pm$ 0.09 & Direct    &   3   \\
ESO 489-G056& 15.70 & 4.99  & rgb & -13.07 & 7.49 $\pm$ 0.10 & -1.35 $\pm$ 0.20 & Direct    &  15   \\
      DDO 53& 14.55 & 3.56  & rgb & -13.37 & 7.62 $\pm$ 0.20 &      \nodata     & Empirical &  1    \\
ESO 444-G084& 15.06 & 4.61  & rgb & -13.56 & 7.45 $\pm$ 0.20 & -1.04 $\pm$ 0.20 & Empirical &  7   \\
    UGC 6541& 14.32 & 3.89  & rgb & -13.71 & 7.82 $\pm$ 0.06 & -1.45 $\pm$ 0.13 & Direct    &  16   \\
       Sex A& 11.86 & 1.32  & cep & -13.93 & 7.54 $\pm$ 0.10 & -1.54 $\pm$ 0.13 & Direct    &  1,17    \\
         WLM& 11.03 & 0.92  & rgb & -13.95 & 7.77 $\pm$ 0.10 & -1.46 $\pm$ 0.05 & Direct    &  5,18,19  \\
       Sex B& 11.85 & 1.36  & rgb & -13.96 & 7.69 $\pm$ 0.15 & -1.46 $\pm$ 0.06 & Direct    &  9,17   \\
    UGC 6456& 14.32 & 4.34  & rgb & -14.03 & 7.73 $\pm$ 0.05 & -1.54 $\pm$ 0.08 & Direct    &  20   \\
     DDO 154& 14.17 & 4.3   & bs  & -14.04 & 7.67 $\pm$ 0.06 & -1.68 $\pm$ 0.13 & Direct    &  12,21 \\
ESO 381-G020& 14.44 & 4.6   &  h  & -14.15 & 7.87 $\pm$ 0.20 & -1.62 $\pm$ 0.20 & Empirical &  7,22 \\
    UGC 9240& 13.10 & 2.79  & rgb & -14.18 & 7.95 $\pm$ 0.03 & -1.60 $\pm$ 0.06 & Direct    &  3,10 \\
     UGC 685& 14.22 & 4.79  & rgb & -14.43 & 8.00 $\pm$ 0.03 & -1.45 $\pm$ 0.08 & Direct    &  3    \\
     UGCA 92& 15.22 & 1.8   & bs  & -14.48 & 7.65 $\pm$ 0.20 &      \nodata     & Empirical &  18   \\
        Ho I& 13.64 & 3.84  & rgb & -14.49 & 7.70 $\pm$ 0.20 &      \nodata     & Empirical &  23   \\
     IC 1613&  9.92 & 0.73  & cep & -14.51 & 7.62 $\pm$ 0.05 & -1.13 $\pm$ 0.18 & Direct    &  7    \\
    UGCA 442& 13.58 & 4.27  & rgb & -14.64 & 7.72 $\pm$ 0.03 & -1.41 $\pm$ 0.02 & Direct    & 24,25 \\
     IC 4662& 11.74 & 2.0   &   h & -15.07 & 8.17 $\pm$ 0.04 & -1.50 $\pm$ 0.05 & Direct    & 26,27,28\\
ESO 383-G087& 11.03 & 1.5   &  h  & -15.16 & 8.19 $\pm$ 0.06 & -1.37 $\pm$ 0.08 & Direct    &   7   \\
    NGC 6822&  9.32 &  0.5  & cep & -15.20 & 8.11 $\pm$ 0.05 & -1.60 $\pm$ 0.10 & Direct    & 1,29  \\
     DDO 168& 12.97 & 4.33  & rgb & -15.28 & 7.50 $\pm$ 0.20 &      \nodata     & Empirical &  10   \\
    NGC 3109& 10.39 & 1.33  & rgb & -15.52 & 7.73 $\pm$ 0.33 & -1.32 $\pm$ 0.20 & Direct    &  7,30 \\
       IC 10& 12.20 & 0.66  & cep & -15.55 & 8.19 $\pm$ 0.14 &      \nodata     & Direct    &  30   \\
ESO 245-G005& 12.73 & 4.43  & rgb & -15.57 & 7.70 $\pm$ 0.10 & -1.27 $\pm$ 0.10 & Direct    &  24,28 \\
     IC 5152& 11.06 & 2.07  & rgb & -15.63 & 7.92 $\pm$ 0.07 & -1.05 $\pm$ 0.12 & Direct    &  7    \\
    NGC 2915& 13.20 & 3.78  & rgb & -15.88 & 8.30 $\pm$ 0.10 & -1.3  $\pm$ 0.1  & Empirical &  7    \\
    NGC 5264& 12.60 & 4.53  & rgb & -15.90 & 8.66 $\pm$ 0.20 & -0.57 $\pm$ 0.20 & Empirical &  7    \\
    NGC 2366& 11.68 & 3.19  & rgb & -16.00 & 7.91 $\pm$ 0.05 &      \nodata     & Direct    &  31   \\
         SMC&  2.75 & 0.06  & cep & -16.31 & 8.13 $\pm$ 0.10 & -1.58 $\pm$ 0.15 & Direct    &  37    \\
    NGC 5408& 12.21 & 4.81  & rgb & -16.50 & 8.00 $\pm$ 0.03 & -1.46 $\pm$ 0.05 & Direct    &  26,32 \\
     NGC 625& 11.59 & 4.07  & rgb & -16.53 & 8.10 $\pm$ 0.10 & -1.25 $\pm$ 0.03 & Direct    &  25    \\
    NGC 1560& 11.90 & 3.45  & rgb & -16.60 & 8.10 $\pm$ 0.20 &      \nodata     & Empirical &  30    \\
       Ho II& 11.09 & 3.39  & rgb & -16.70 & 7.92 $\pm$ 0.10 &      \nodata     & Direct    &  30,31 \\
    NGC 4214& 10.24 & 2.94  & rgb & -17.19 & 8.25 $\pm$ 0.10 & -1.30 $\pm$ 0.15 & Direct    &  31,33 \\
     IC 2574& 10.84 & 4.02  & rgb & -17.34 & 8.09 $\pm$ 0.07 & -1.52 $\pm$ 0.13 & Direct    &  23,31 \\
    NGC 5253& 10.87 & 4.00  & cep & -17.38 & 8.15 $\pm$ 0.10 & -0.84 $\pm$ 0.10 & Direct    &  34   \\
      NGC 55&  8.84 & 1.8   & tf  & -17.50 & 8.35 $\pm$ 0.10 &      \nodata     & Empirical &  35   \\
     UGCA 86& 13.50 & 2.65  &  bs & -17.68 & 7.80 $\pm$ 0.20 &      \nodata     & Empirical &  18   \\
     NGC 300&  8.95 & 2.15  & cep & -17.77 & 8.73 $\pm$ 0.04 &      \nodata     & Empirical &  35   \\
    NGC 4395& 10.61 & 4.61  & rgb & -17.78 & 8.45 $\pm$ 0.20 & -1.52 $\pm$ 0.20 & Empirical &  36   \\
         LMC&  0.90 & 0.05  & cep & -17.91 & 8.37 $\pm$ 0.22 & -1.30 $\pm$ 0.20 & Direct    &  37   \\
\enddata
\tablecomments{Optical parameters and distances are taken from the compilation of 
Karachentsev et al.\ 2004.  References for oxygen and nitrogen abundances: 
0 - present paper;
1 - Skillman et al.\ (1989a);
2 - van Zee (2000);
3 - van Zee \& Haynes (2005);
4 - Skillman et al.\ (1997);
5 - Skillman et al.\ (1989b);
6 - Saviane et al.\ (2002);
7 - Lee et al.\ (2003a);
8 - Skillman et al.\ (1988);
9 - Moles et al.\ (1990);
10 - Hidalgo-G\'amez \& Olofsson (2002);
11 - Silva et al.\ (2005);
12 - van Zee et al.\ (1997a);
13 - van Zee et al.\ (1997b);
14 - Skillman et al.\ (1994);
15 - R\"onnback \& Bergvall (1995);
16 - Guseva et al.\ (2000);
17 - Kniazev et al.\ (2005);
18 - Hodge \& Miller (1995);
19 - Lee et al.\ (2005a);
20 - Izotov et al.\ (1997);
21 - Kennicutt \& Skillman (2001);
22 - Webster et al.\ (1983);
23 - Miller \& Hodge (1996);
24 - Miller (1996);
25 - Skillman et al.\ (2003);
26 - Stasi\'nska et al.\ (1986);
27 - Heydari-Malayeri et al.\ (1990);
28 - Hidalgo-G\'amez et al.\ (2001a);
29 - Hidalgo-G\'amez et al.\ (2001b);
30 - Lee et al.\ (2003b);
31 - Masegosa et al.\ (1991);
32 - Masegosa et al.\ (1994);
33 - Kobulnicky \& Skillman (1996);
34 - Kobulnicky et al.\ (1997);
35 - Zaritsky et al.\ (1994);
36 - van Zee et al.\ (1998);
37 - Russell \& Dopita (1990).
}
\end{deluxetable}

\begin{figure}
\epsscale{0.8}
\plotone{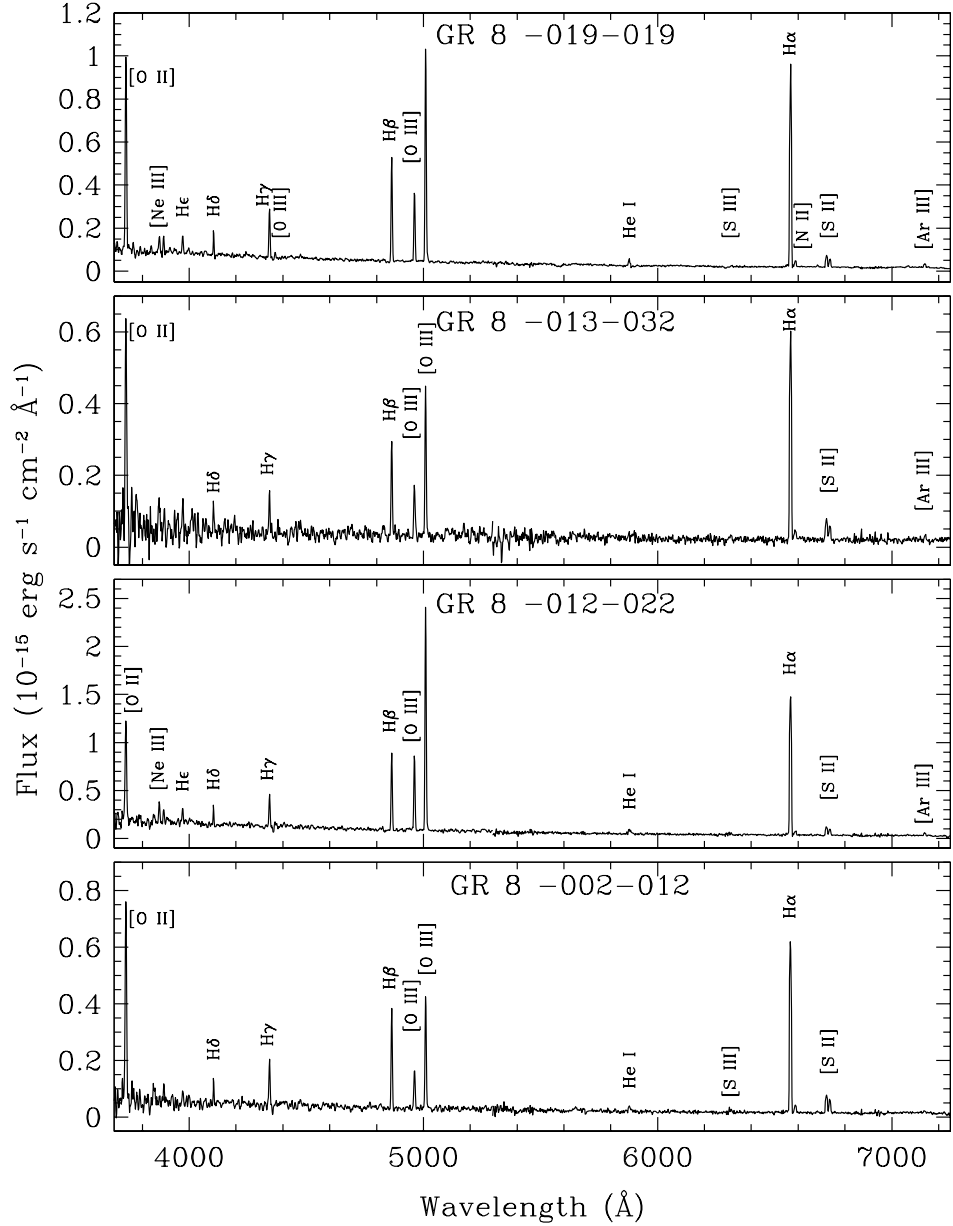}
\caption[]{
Optical spectra of HII regions in GR 8.  The low metallicity nature
of GR 8 is clearly demonstrated by the weak [N~II] and [S~II] lines.
\label{fig:GR8} }
\end{figure}
\setcounter{figure}{2}
\begin{figure}
\epsscale{0.8}
\plotone{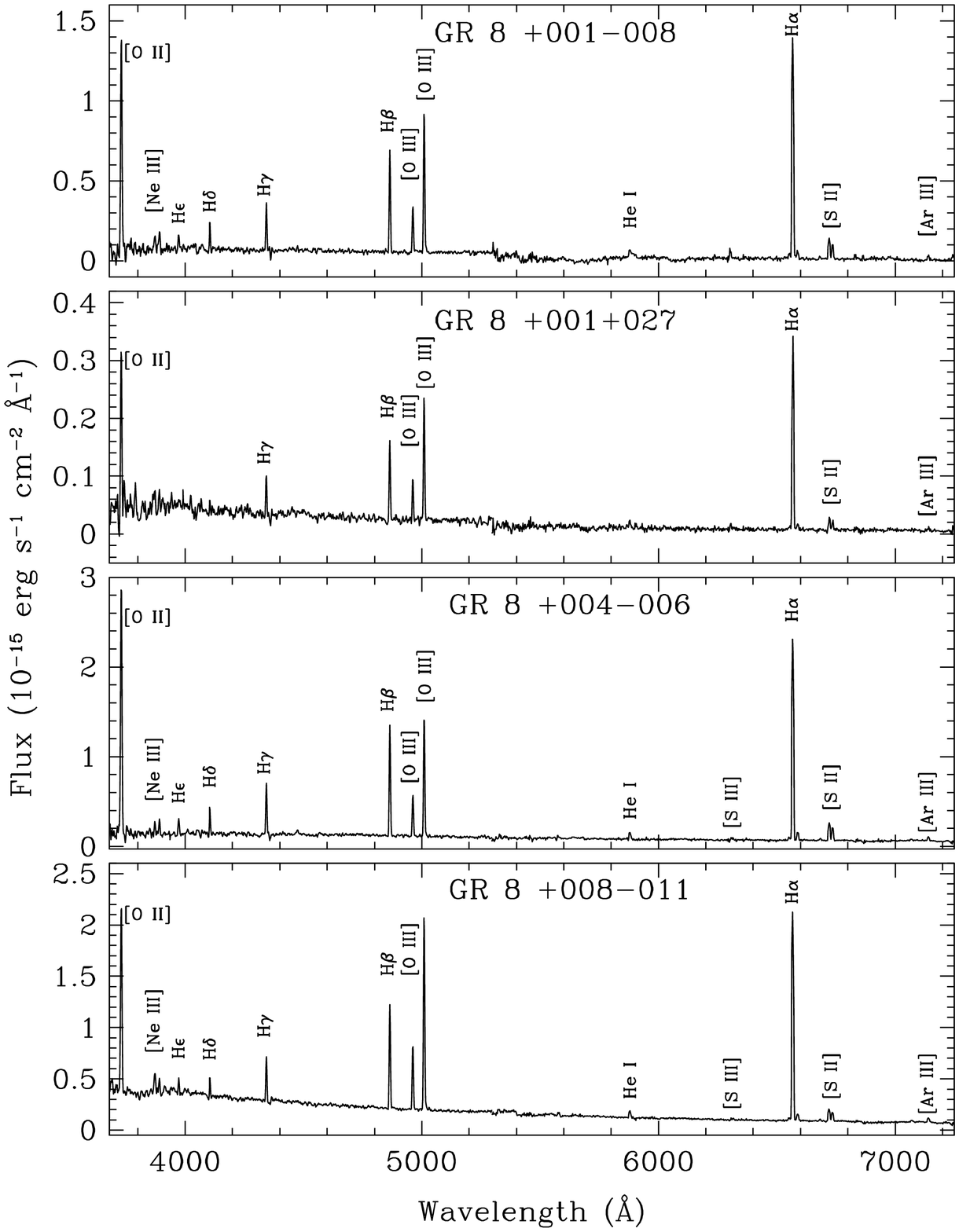}
\caption[]{
continued}
\end{figure}

\begin{figure}
\epsscale{1.0}
\plotone{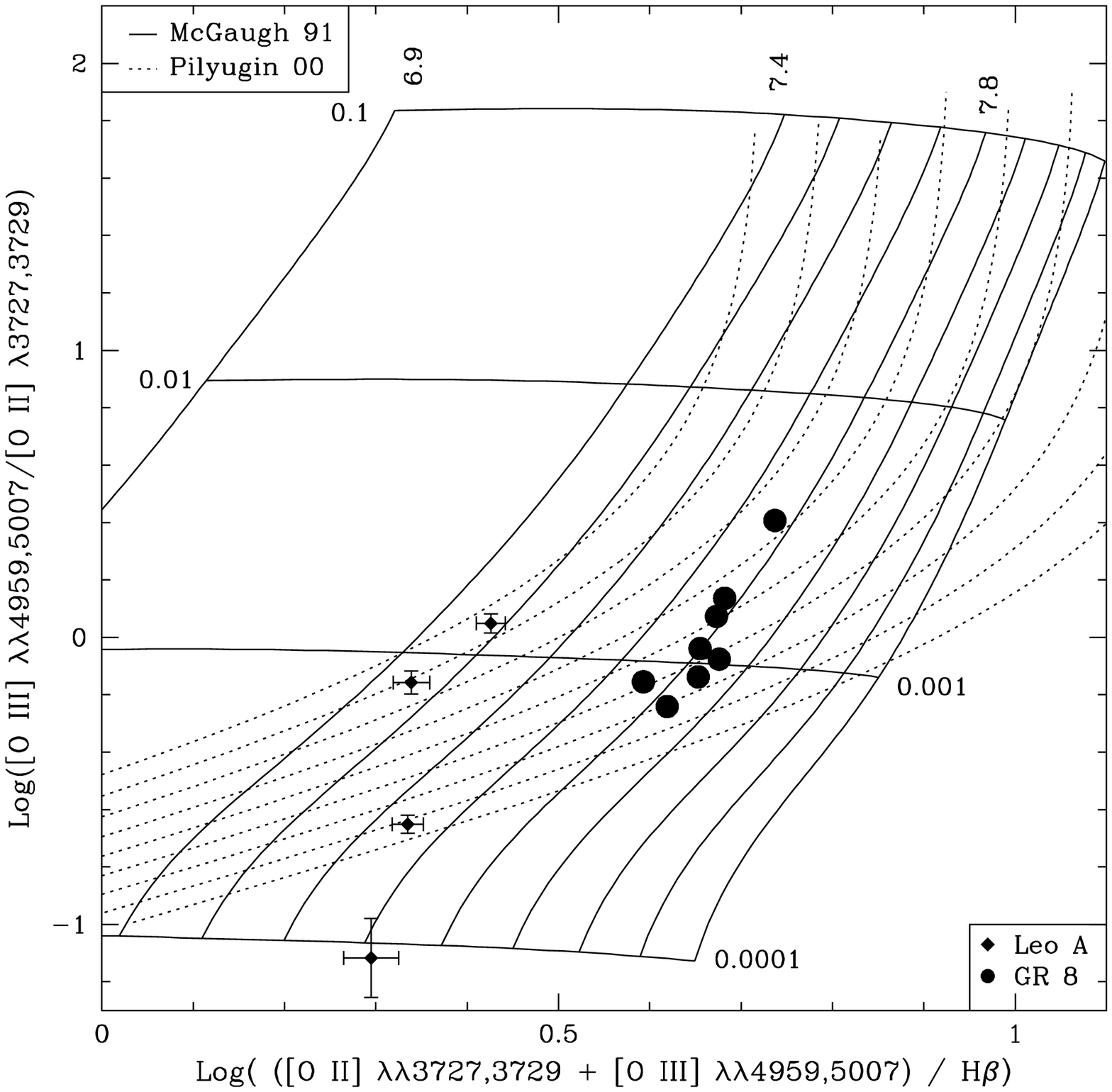}
\caption[]{
Model grid of the lower branch of the R$_{23}$ relation from McGaugh (1991) (solid
lines) for a range of metallicities and ionization parameters. 
 The locations of the HII regions in  Leo~A (diamonds) and GR~8 (circles)
are marked; for GR~8, the error bars are on the order of the
size of the symbols.   Also shown are the empirical metallicity calibration 
relations derived by Pilyugin (2000) for 7.4 $<$ 12 + log(O/H) $<$ 8.2 
(dashed lines, spaced by 0.1 dex).  Note how the HII regions in
GR~8 define a locus similar to the constant metallicity lines from 
McGaugh (1991) but not those defined by the Pilyugin (2000) relations.
Also note that the HII regions in Leo~A do not follow a single value of 
the metallicity for either calibration (see text).
\label{fig:r23} }
\end{figure}

\begin{figure}
\epsscale{0.90}
\plotone{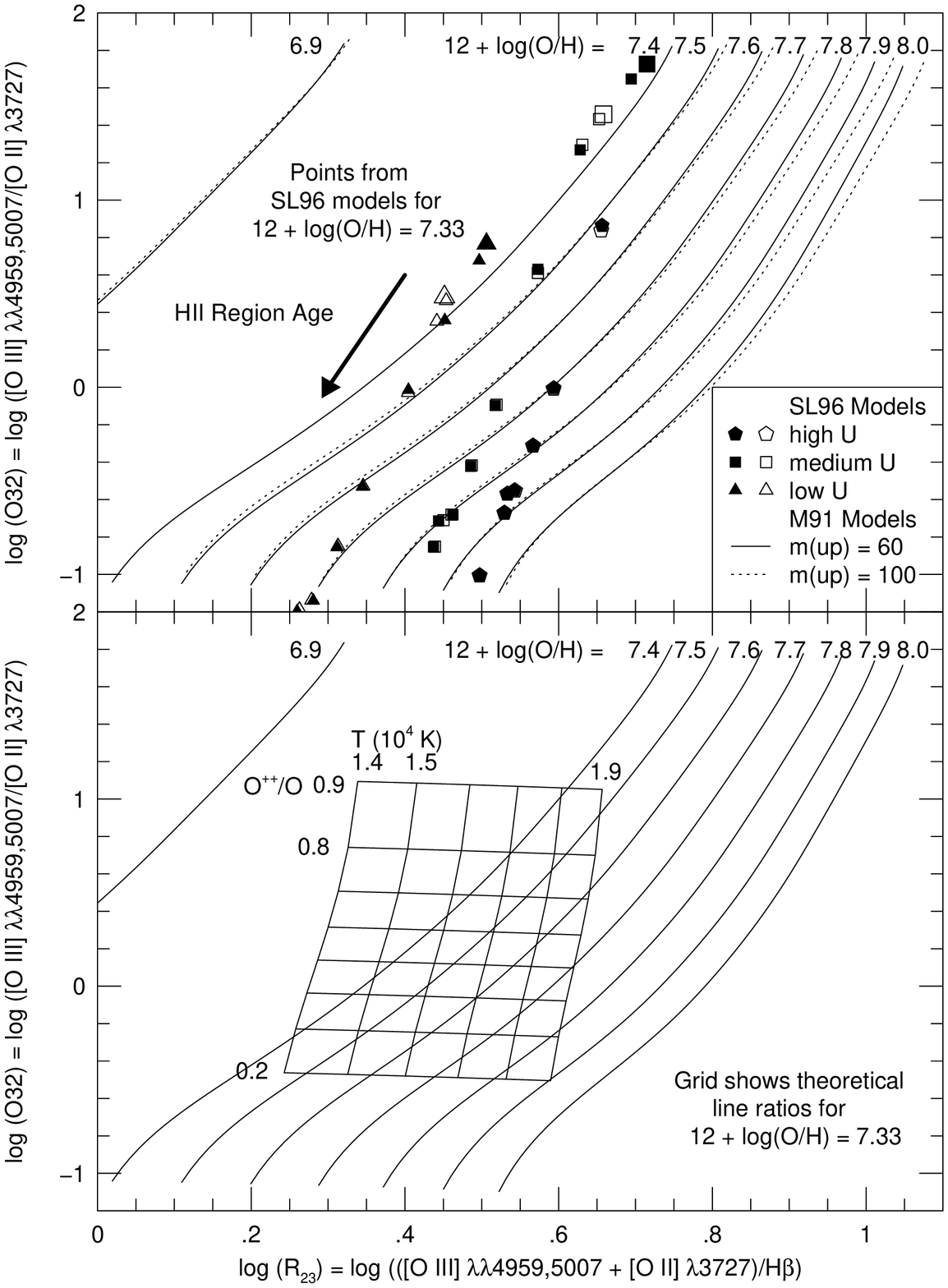}
\caption[]{
(Upper) Model grid of the lower branch of the R$_{23}$ relation from 
McGaugh (1991) (solid lines and dashes).  The results of models of 
HII regions from \citet{SL96} covering ages ranging from 1 - 10 Myr
are shown for 12 $+$ log (O/H) $=$ 7.33.  Large symbols denote 
the value for age = 1 Myr; time steps are 1 Myr intervals.
The models include a range of ionization parameters
and upper limits for the IMF; solid symbols are for 
M(up) = 100 M$_{\odot}$ and open symbols are for M(up) = 50 M$_{\odot}$.
While the models of young HII regions (age $<$ 3 Myr) follow the 
relationships of McGaugh (1991), older HII regions appear to 
show higher oxygen abundance in these diagnostic diagrams. 
(Lower) Same as upper with a grid denoting the positions 
of HII regions for a single abundance of 12 $+$ log (O/H) $=$ 7.33
with a range in electron temperature and ionic fractions.
Clearly a large range in the diagnostic diagram is allowed.
 The empirical abundance
calibrations rely on a very close relationship between temperature
and ionic fraction which is determined by both the spectrum of the
ionizing radiation and the ionization parameter.  Variations from
this relationship (introduced by either truncating the IMF or 
aging the ionizing cluster) result in large departures from the 
oxygen abundance loci in the diagnostic diagram.
\label{fig:grid} }
\end{figure}

\begin{figure}
\epsscale{1.0}
\plotone{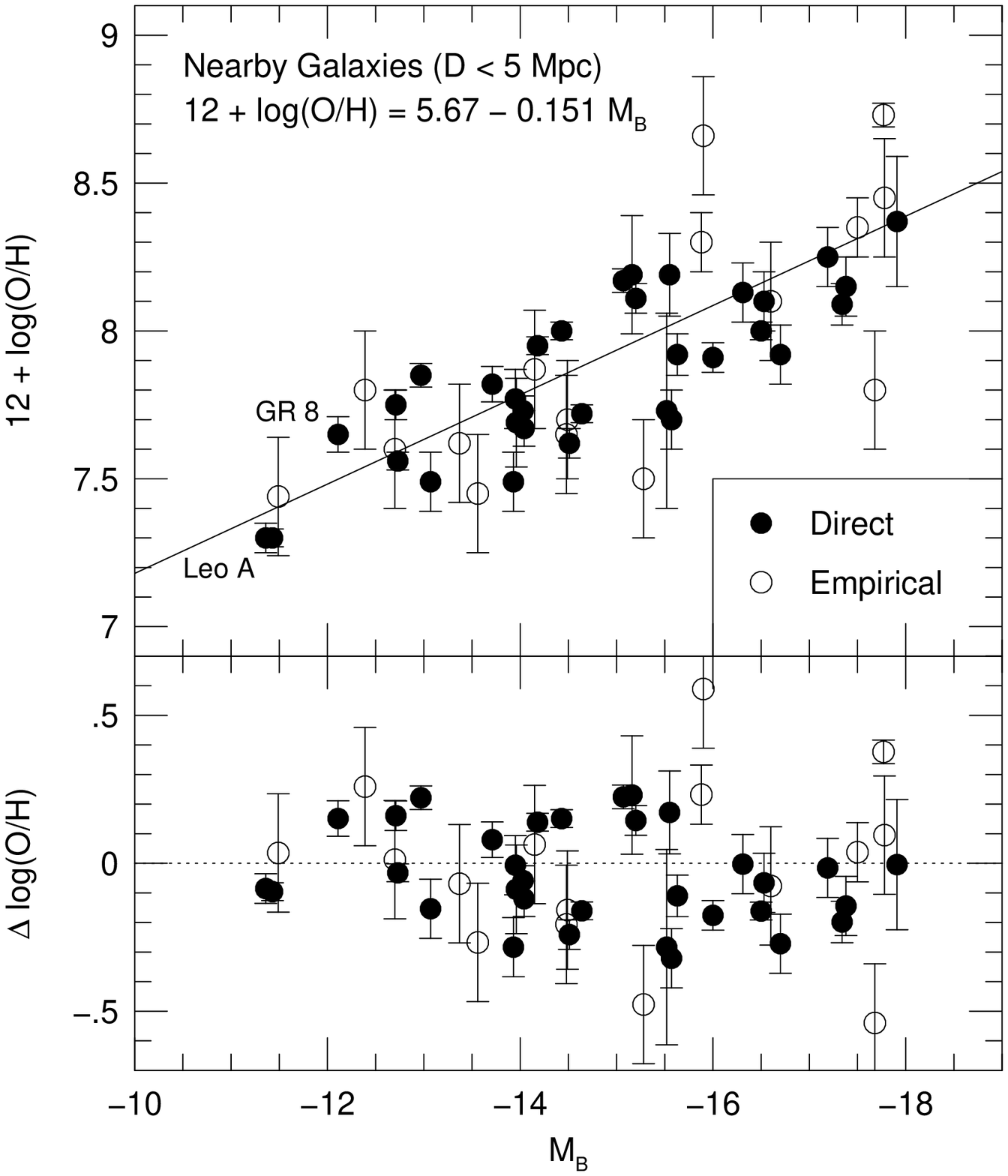}
\caption[]{
(Upper) The metallicity-luminosity relation for nearby (D $<$ 5 Mpc) gas-rich
galaxies fainter than M$_B$ of --18.  The new measurements of the
oxygen abundances of Leo~A and GR~8 provide valuable data at the low
luminosity end of this relationship.  Filled symbols indicate that the
oxygen abundance was derived via a ``direct'' measurement, while 
open symbols indicate that the oxygen abundance was derived via 
the bright-line method.  The solid line represents a weighted least
squares fit to all the points.
(Lower) Deviations from the average relationship between log(O/H) and M$_B$.
Note that the dispersions are similar for the direct and bright-line
abundances and for the low and high luminosity galaxies (see text).
\label{fig:metlum} }
\end{figure}

\begin{figure}
\plotone{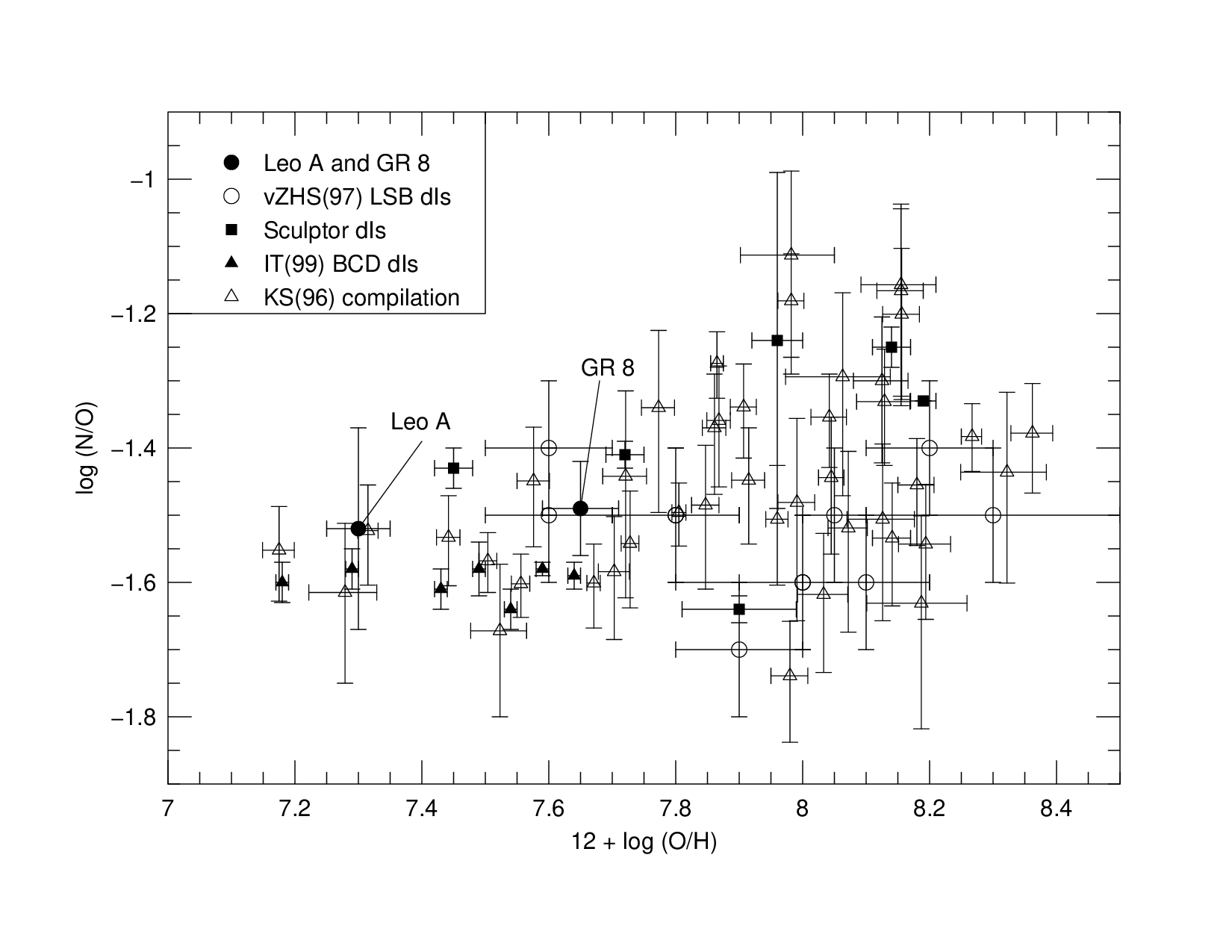}
\caption[vanzee.fig08.ps]{
Comparison of the N/O and O/H in Leo~A and GR~8 (filled circles)
with other star forming dwarf galaxies from the literature.  
The empty circular symbols represent data for low surface brightness dwarf
irregular galaxies from van Zee et al.\ (1997a). 
Four Sculptor dwarf irregular galaxies with direct abundance measurements 
are shown with filled squares \citep{SCM03}.
The filled triangles represent the low metallicity blue compact dwarf 
galaxies from Izotov \& Thuan (1999).  
The collection of dwarf irregular galaxies and H~II galaxies
assembled by Kobulnicky \& Skillman (1996; see their Table 5 and Figure 15
for identification of individual points) are represented by open triangles.
Only galaxies without WR emission features
and errors in log (N/O) less than 0.2 have been plotted.
\label{fig:no} }
\end{figure}
\end{document}